\documentclass[aps,scriptaddress, showkeys,nofootinbib,reprint]{revtex4-1}
	\usepackage{hyperref}
	\usepackage{cancel} 
	\usepackage{amsmath,amssymb} 
	\usepackage{makeidx} 
	\usepackage{amsfonts}
	\usepackage[ansinew]{inputenc}
	\usepackage[usenames,dvipsnames]{pstricks}    
	\usepackage{subfigure}
	\usepackage{epsfig}
	\usepackage{pst-grad}
	\usepackage{pst-plot} 
	\usepackage{comment}
	\usepackage{pstool}
\makeindex
\newcommand{\beq}{\begin{equation}}
\newcommand{\eeq}{\end{equation}}
 
\newcommand{\bea}{\begin{eqnarray}}
\newcommand{\ea}{\end{eqnarray}}
\newcommand{\barr}{\begin{array}}
\newcommand{\earr}{\end{array}}
\newcommand{\lb}{{\langle}}
\newcommand{\rb}{{\rangle}}

\newcommand{\calo}{{\cal O}}

\begin{document}

\title{Warm Ekpyrosis}

\author{Aaron M. Levy}
\email{aaronml@princeton.edu}
\affiliation{Department of Physics, Princeton University, Princeton, NJ 08544, USA}
\author{Gustavo J. Turiaci}
\email{turiaci@princeton.edu}
\affiliation{Department of Physics, Princeton University, Princeton, NJ 08544, USA}

\begin{abstract}
We propose a mechanism to generate a nearly scale-invariant spectrum of adiabatic scalar perturbations about a stable, ekpyrotic background. The key ingredient is a coupling between a single ekpyrotic field and a perfect fluid of ultra-relativistic matter. This coupling introduces a friction term into the equation of motion for the field, opposing the Hubble anti-friction, which can be chosen such that an exactly scale-invariant (or nearly scale-invariant) spectrum of adiabatic density perturbations is continuously produced throughout the ekpyrotic phase. This mechanism eliminates the need for a second (entropic) scalar field and hence any need for introducing a second phase for converting entropic into curvature fluctuations.  It also reduces the constraints on the equation of state during the ekpyrotic phase and, thereby, the need for parametric fine-tuning. 
\end{abstract}

\date{\today}

\maketitle
\section{Introduction}

Inflation is one proposed mechanism for smoothing and flattening the universe while simultaneously stretching primordial quantum fluctuations to superhorizon scales \cite{PhysRevD.23.347,Albrecht:1982wi,Linde:1981mu}.  Ekpyrotic contraction is another \cite{Khoury:2001bz}.  In both cases, the fluctuations evolve into the seeds of large-scale structure and imprint observable anisotropies onto the cosmic microwave background.  The \emph{Wilkinson Microwave Anisotropy Probe} (WMAP) \cite{Komatsu:2008hk}, the \emph{Planck} satellite \cite{Ade:2015lrj,Ade:2015ava}, the \emph{Atacama Cosmology Telescope} (ACT) \cite{Sievers:2013ica} and other experiments indicate that this primordial density fluctuation spectrum is adiabatic and nearly scale-invariant with nearly Gaussian statistics. Although inflation can generate such perturbations, it requires rare initial conditions  \cite{Penrose:1988mg,Gibbons:2006pa,Berezhiani:2015ola}  and results in a multiverse of outcomes \cite{1983veu..conf..251S, Vilenkin:1983xq, Guth:2000ka, Guth:2013epa, Guth:2013sya,  Linde:2014nna}. Given that ekpyrosis avoids these pathologies, it is natural to ask whether it can generate the same spectrum of perturbations. 

In ekpyrotic universes, smoothing contraction occurs because the energy density of a scalar field, $\phi$, with equation of state $\epsilon_{\phi} > 3$ (where $\epsilon_\phi = 3(p_\phi + \rho_\phi)/\rho_\phi$ with $p_\phi$ being the pressure and $\rho_\phi$ the energy density of the scalar field) grows to dominate all other forms of energy, including inhomogeneities, anisotropy, and spatial curvature. Meanwhile, due to the slow contraction, fluctuation modes shrink more slowly than the Hubble radius, so that quantum fluctuations escape to cosmological scales. 

The earliest models of ekpyrosis involved a single, minimally coupled scalar field with a steep, negative exponential potential. After some debate, it was shown that these models cannot produce the observed scale-invariant, adiabatic spectrum because the comoving curvature perturbation acquires a strong blue tilt \cite{Brandenberger:2001bs, Lyth:2001pf, Hwang:2001ga, Hwang:2002ks,Creminelli:2004jg}. However, it was noticed in Ref.~\cite{Finelli:2002we} that when a second scalar field is added-- also with a steep, negative potential-- there exists a background solution along the potential energy surface whose entropic perturbations acquire a scale-invariant spectrum. After the ekpyrotic smoothing phase, it was argued in Ref.~\cite{Lehners:2007ac}, the entropic perturbations will convert into a scale-invariant spectrum of adiabatic perturbations if the background solution undergoes a bend in field-space. This two-step process, first of generating scale-invariant entropic perturbations and then of converting them to adiabatic perturbations, has been dubbed the ``entropic mechanism'' \cite{Buchbinder:2007ad, DiMarco:2002eb}. The first models making use of the entropic mechanism require finely-tuned initial conditions because the background solution is unstable to small perturbations \cite{Tolley:2007nq,Koyama:2007ag,Koyama:2007mg,Buchbinder:2007tw,Hinterbichler:2011qk}. 

More recent two-field models have cured this instability by introducing non-canonical kinetic terms \cite{2013PhLB..724..192L,Li:2014yla,Ijjas:2014fja,Levy:2015awa}.  Such terms provide friction in the equation of motion for the non-canonically coupled field \cite{Notari:2002yc}. This friction has two effects:  1)  it damps the background evolution for the non-canonically coupled field, thereby making it the entropy direction in field-space and 2) it alters the spectrum of perturbations in this direction:  scale-invariant entropic spectra are produced even though the entropy field has no potential. These newer models have the attractive features that they generate no detectable spectrum of primordial gravitational waves (the ratio of the tensor-perturbation amplitude to the scale-perturbation amplitude, $r\approx 0$) and zero non-Gaussianity during the ekpyrotic contraction phase; only a small amount of local non-Gaussianity ($f_{\rm NL}=\mathcal{O}(1)$) is generated by the conversion process \cite{Boyle:2003km,Fertig:2013kwa,Ijjas:2014fja}. These models also impose less stringent constraints on the equation of state parameter of the universe and hence require less fine-tuning of parameters than actions with canonical kinetic terms.

In all of these models, during the slow contracting phase, the ekpyrotic fields are assumed to have no direct interaction with any other fields.  They simply traverse their potential energy surface in a supercooled universe, and only after the ekpyrotic phase is the universe assumed to reheat, either through some coupling to Standard Model particles or through stringy, higher dimensional effects \cite{Buchbinder:2007ad}. In this work, we consider a single, ekpyrotic field coupled to a perfect fluid of ultra-relativistic matter (\emph{e.g.,} radiation) in thermal equilibrium.  As the field falls down its steep, negative potential, it decays continuously into lighter fields which are thermally excited, thus generating a dissipative friction term in its equation of motion. As in the non-canonical, two-field models discussed above, this friction term allows a scale-invariant spectrum to be produced. In contrast to the non-canonical models, the scale-invariant spectrum is immediately adiabatic; no conversion is necessary. 

To describe the interaction between the fluid and scalar field, we strive for generality, leaving the details of specific microphysical model-building for future work.  Therefore, we work at the level of the equations of motion, adding generic dissipative and noise terms.  As we will show, if the dissipation is too strong, the radiation fluid dominates the energy density of the universe; if it is too weak, the scalar field dominates.  Hence, our results require that the dissipation coefficient evolve in fixed proportion to the Hubble parameter; this is the main source of fine-tuning (although this tuning can be relaxed somewhat by changing the details of the interaction).   

The idea of particle production during a cosmological smoothing and flattening phase has been investigated previously in models referred to as warm inflation, where thermal fluctuations sourced by radiation-induced noise were shown to dominate over vacuum fluctuations \cite{Berera:1995wh,Berera:1995ie,Berera:1996nv,Berera:2006xq,Berera:2008ar,BasteroGil:2009ec}. Similar effects appear in models such as trapped inflation \cite{Green:2009ds}. In contracting universes, however, the thermal fluctuations are suppressed on the largest scales, and the density perturbations are dominated by vacuum fluctuations.  The reason is that contracting universes grow hotter with time, so that longer modes cross the horizon at lower temperature with correspondingly smaller thermal fluctuations. 

This paper is organized as follows. In Sec.~\ref{backgroundsec}, we solve and analyze the background evolution, showing the appearance of a new family of attractors introduced by the interaction between the ekpyrotic field and the radiation fluid. In Sec.~\ref{pertsec}, we compute the power spectrum for the comoving curvature perturbation by studying scalar perturbations to linear order.  This results in a Langevin-like equation that we solve using Green's function techniques. We find that the thermal contribution to the power spectrum is subleading to the vacuum contribution over the observable modes, and we show how to fix the parameters of the model to obtain scale invariance. In Sec.~\ref{conc}, we discuss implications of our results and directions for future work.

\section{Background}\label{backgroundsec}
\begin{figure}
\includegraphics[width=0.45\textwidth]{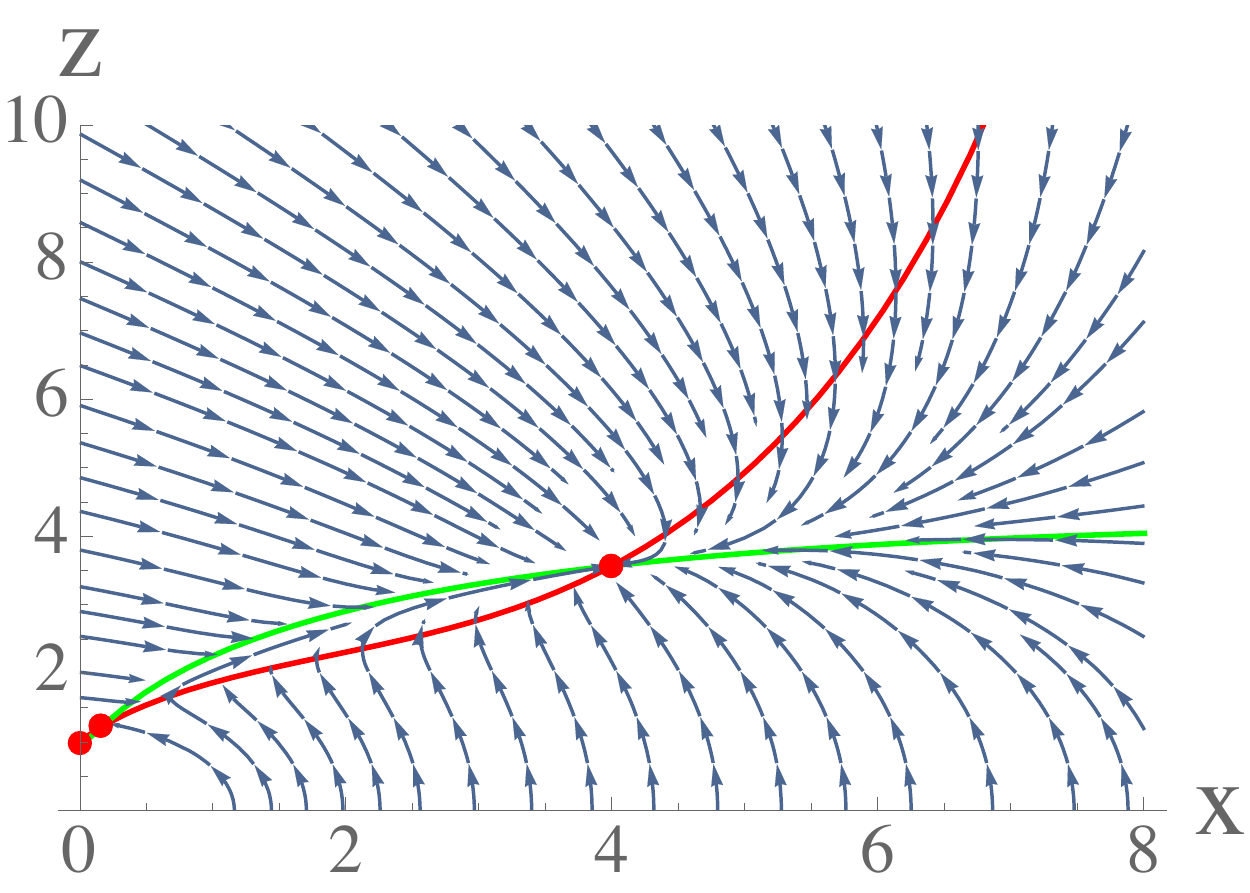}
\caption{\label{fixedpointcurve} This streamplot shows that the warm ekpyrotic background solution is an attractor for a wide range of initial conditions. For illustration, we have chosen parameter values $(c,\gamma)=(15,-56.8)$. Any set of initial conditions for $\phi$, $\dot\phi$, $\dot\rho$, and $a$ corresponds to a particular point in this plane $(x,z)~\equiv~\left(\dot\phi/(\sqrt{6}H),-\sqrt{\rho_r}/(\sqrt{3}H)\right)$. The background solution follows the blue arrows originating at this point. The red and green curves are included simply to guide the eye:  they are nullclines, where $dx/d\ln a=0$ (red) and $dz/d\ln a=0$ (green).  The intersections of the nullclines are shown as red dots. These so called ``fixed-point, scaling solutions'' are special because the blue streamlines vanish here. If the background solution starts at one of these points, it stays there. The rightmost such point corresponds to $(x_0,z_0)$ defined in Eqs.~\eqref{pinkx1} and \eqref{pinkz1}.  Clearly, it is an attractor for a wide range of initial conditions.  The analysis below shows that the comoving curvature perturbation generated by this solution acquires a scale-invariant spectrum on large scales. 
}
\end{figure}

In this section, we derive an explicit solution for the background cosmology.  The main results of this section are Eqs.~\eqref{bgsolutions} and Fig.~\ref{fixedpointcurve}.

We employ reduced Planck units in which $8\pi G_N=k_B=\hbar=c_{L}=1$ where $G_N$ is Newton's gravitational constant, $k_B$ is Boltzmann's constant, $\hbar$ is the reduced Planck's constant, and $c_{L}$ is the speed of light. We use the metric signature $(-,+,+,+).$  Commas denote ordinary derivatives, and semicolons denote covariant derivatives.  

We consider a contracting universe populated with a radiation fluid and a minimally coupled scalar field obeying Einstein's equations, 
\begin{equation}
\label{ee}
G_{ab}=T_{ab}.
\end{equation}  
Here, $G_{ab}$ is the Einstein tensor, and $T_{ab}=T^{(r)}_{ab}+T^{(\phi)}_{ab}$ is the total energy-momentum tensor which has been decomposed into a term describing the radiation fluid, denoted by the superscript $(r)$, and a term describing the scalar field, denoted by the superscript ($\phi$). The radiation fluid is characterized by a four-velocity, $u^a$, an energy density, $\rho_r,$ and a pressure, $p_r$, so that its energy-momentum tensor is given by
\begin{equation}
T^{(r)}_{ab}=(\rho_r+p_r)u_a u_b+p_r g_{ab}.
\end{equation}
For simplicity, we take $p_r=\rho_r/3$, although this is not central to our results. The scalar field is characterized by a potential energy density, $V(\phi)$, so that its energy-momentum tensor is given by
\begin{equation}
T^{(\phi)}_{ab}=\phi_{,a}\phi_{,b}-\left(\frac{1}{2}\phi^{;c}\phi_{;c}+V(\phi)\right)g_{ab}.
\end{equation}
For convenience, we take the negative, exponential form, $V(\phi)=V_{0}e^{-c\phi},$ where $V_{0}<0$ and $c>0$. The interaction between the radiation fluid and the scalar field is described by a flux term,  $Q_a\equiv-\Gamma u^b \phi_{,b}\phi_{,a}$, satisfying
\begin{eqnarray}
\label{cons}
T^{(r)b}_{\,\,\,\,\,\,\,\,\,\,\,a;b}=-T^{(\phi)b}_{\,\,\,\,\,\,\,\,\,\,\,a;b}=Q_a.
\end{eqnarray}

In a spatially flat, Friedmann-Robertson-Walker spacetime, the background metric takes the form 
\begin{eqnarray}
ds^2&=&a^2(\tau)\left(-d\tau^2+\delta_{ij}dx^idx^j\right)\\
&=&-dt^2+a^2(t)\delta_{ij}dx^idx^j,
\end{eqnarray}
where $a$ is the scale factor, $t<0$ is cosmic time, and $\tau<0$ is conformal time defined by $d\tau\equiv dt/a.$ At background, Eqs.~\eqref{cons} gives two equations (from the $t$-component)
\begin{eqnarray}
\label{phi}
\ddot{\phi}+3H\dot{\phi}+V,_{\phi}&=&-\Gamma\dot{\phi},\label{kgf}\\
\label{rad}
\dot{\rho}_r+4H\rho_r&=&\Gamma\dot{\phi}^2\label{eaop},
\end{eqnarray}
where overdots represent derivatives with respect to cosmic time and $H\equiv \dot{a}/a<0$ is the Hubble parameter. The flux term, proportional to $\Gamma$, describes the decay of the $\phi$-field into the particles comprising the radiation fluid.  It appears in two places: on the right side of Eq.~\eqref{eaop}, it sources the energy density of the radiation, ensuring that $\rho_r$ is not rapidly outstripped by the energy density in the ekpyrotic field, $\phi$; but most important, in Eq.~\eqref{kgf} it manifests as a dissipative friction term for $\phi$.  As we will see, this friction term is critical for the production of a scale-invariant spectrum. 

As it stands, Eq.~\eqref{phi} for the scalar field is incomplete. It is well understood in the context of classical and quantum theory that whenever a process generates an effective dissipative interaction, it also generates fluctuations that can be described by a stochastic noise source, $\Xi$,  with zero mean \cite{calzetta2008nonequilibrium}.  Thus, Eq.~\eqref{phi} should read
\beq
\ddot{\phi}+3H\dot{\phi}+V,_{\phi}=-\Gamma\dot{\phi}+\Xi.
\eeq
If the microphysical process responsible for this noise, $\Xi,$ is in thermal equilibrium at some temperature, $T$, then the Fluctuation-Dissipation Theorem relates the dissipation it induces, $\Gamma$, to its correlation function via
\beq
\lb \Xi(x,\tau)\Xi(x',\tau')\rb= 2 \Gamma T \delta^{(3)}(x-x')\delta(\tau'-\tau), \label{noiseterms}
\eeq 
where angular brackets denote ensemble averaging. If this process is not in thermal equilibrium, then its correlation, $\lb \Xi\Xi\rb$, can depend more generally on $(x,\tau)$ and $(x',\tau')$. As discussed in the introduction, this noise term is critical in warm inflation because it significantly enhances the power spectrum of scalar perturbations relative to the vacuum result.  In the contracting models considered here, the opposite is true:  as we will show, the noise, $\Xi$, is completely irrelevant to the power spectrum of the comoving curvature perturbation.   Moreover, since it has zero mean, $\lb \Xi\rb$, it is irrelevant to the background dynamics as well and will be omitted in the reminder of this section.

To find the background dynamics, Eqs.~\eqref{phi} and \eqref{rad} must be solved subject to the Friedmann constraint (from the $t$-$t$ component of Eq.~\eqref{ee}),  
\begin{equation}
\label{fman}
H^2=\frac{1}{3}\left(\frac{1}{2}\dot{\phi}^2+V+\rho_r\right).
\end{equation}
To this end, it proves useful to introduce the dimensionless ``$\Omega$-variables,'' (or more properly their square roots) 
\beq
(x,y,z)\equiv\left(\frac{\dot{\phi}}{\sqrt{6}H},-\frac{\sqrt{|V|}}{\sqrt{3}H},-\frac{\sqrt{\rho_r}}{\sqrt{3}H}\right),\label{xyz}
\eeq   
characterizing respectively the fractional kinetic energy density in the scalar field, the fractional potential energy density in the scalar field, and the fractional energy density in the radiation fluid.  In terms of these variables, Eq.~\eqref{fman} can be rewritten as $y=\sqrt{x^2+z^2-1},$ where we have taken the positive root since $H<0$ in a contracting universe.  The equation of state of the universe takes the simple form 
\begin{equation}
\label{eos}
\epsilon\equiv -\dot{H}/H^2=3x^2+2z^2,
\end{equation}
as can be obtained by differentiating Eq.~\eqref{fman} and substituting Eqs.~\eqref{phi} and \eqref{rad}. Therefore, ekpyrosis occurs whenever $3x^2+2z^2>3$. Meanwhile, Eqs.~\eqref{phi} and \eqref{rad} can be rewritten as
{\small 
\begin{eqnarray}
\frac{dx}{d\ln a}&=&3(x^2+z^2-1)\left(x-\frac{c}{\sqrt{6}}\right)-x\left(z^2+\frac{\Gamma}{H}\right) \label{x},\\
\frac{dz}{d\ln a}&=&(3x^2+2z^2-2)z+\frac{\Gamma}{H}\frac{x^2}{z} \label{z}.
\end{eqnarray}}
Note that $\Gamma$ appears only in the ratio $\gamma\equiv\Gamma/H<0.$ It is at this point that the fine-tuning enters:  we assume in this work that $\gamma$ is a constant, independent of time. To motivate this assumption, note that if $|\gamma|$ grows rapidly, the universe becomes dominated by radiation only, and if it shrinks rapidly, the universe becomes dominated by the ekpyrotic field only.  It is only when $\gamma$ is roughly constant that these two components coexist. Therefore, we assume it in what follows, and merely observe in passing that if $\Gamma\propto \sqrt{\rho_\phi}$, where $\rho_\phi\equiv \frac{1}{2}\dot{\phi}^2+V$, then $\gamma$ is constant along the solution of interest, and our assumption is justified.  With this assumption, Eqs.~\eqref{x} and \eqref{z} admit a fixed-point, scaling solution at $(x_0,z_0)$ with
\begin{eqnarray}
x_0&\equiv&\big[\big(24 \left(c^2+4\right)\gamma+9 \left(c^2-4\right)^2+16\gamma^2\big)^{1/2}\nonumber\\
&&+3 c^2+4\gamma+12\big](6 \sqrt{6} c)^{-1}\label{pinkx1},\\
z_0&\equiv&[(\gamma +3)(\sqrt{9 c^4+24 (\gamma -3) c^2+16 (\gamma+3)^2}\nonumber\\
&&+4 \gamma +12)-3 c^2 (\gamma-3)]^{1/2}(3 \sqrt{2}c)^{-1}.\label{pinkz1}
\end{eqnarray}
As a consistency check, note that in the absence of radiation, \emph{i.e.,} when $\gamma=0,$ this solution reduces to $(x,z)=(c/\sqrt{6},0)$, which reproduces ordinary, single-field ekpyrosis (when $c>\sqrt{6}$). 

To summarize, Eqs.~\eqref{pinkx1} and \eqref{pinkz1} describe a cosmological background whose evolution is given by 
\begin{equation}
\begin{aligned}[c]
a&=(t/t_{\rm e})^{\frac{1}{\epsilon}}\\
H&\equiv \dot a /a=\frac{1}{\epsilon t}\\
\phi&=\phi_{\rm e}+\frac{\sqrt{6}x_0}{\epsilon}\ln\left(t/t_{\rm e}\right)\\
\rho_r &=\frac{3z_0^2}{\epsilon^2 t^2}
\end{aligned}
\qquad
\begin{aligned}[c]
a&=(\tau/\tau_{\rm e})^{\frac{1}{\epsilon-1}}\\
\mathcal{H}&\equiv a'/a=\frac{1}{ (\epsilon -1) \tau }\\
\phi&=\phi_{\rm e}+ \frac{\sqrt{6}x_0}{\epsilon-1}\ln\left(\tau/\tau_{\rm e}\right)\\
\rho_r &=\frac{3 z_0^2 \left(\tau/\tau_{\rm e}\right)^{-2 \epsilon/(\epsilon -1)}}{\tau_{\rm e}^2 (\epsilon -1)^2}\label{bgsolutions},
\end{aligned}
\end{equation}
%
%
where $'\equiv d/d\tau$ and we have normalized the scale factor to unity when ekpyrosis ends at some time $t_{\rm e}<0$.
For convenience, we have included the results in conformal time and defined $\tau_{\rm e}\equiv\epsilon(\epsilon-1)^{-1}t_{\rm e}$ and 
\beq
\label{phi0}
\phi_{\rm e}\equiv \frac{1}{c} \ln \left(-\frac{ V_0\tau_{\rm e}^2(\epsilon -1)^2}{3 \left(x_0^2+z_0^2-1\right)}\right).
\eeq
These dynamics are pictured in Fig.~\ref{fixedpointcurve}, which shows that this solution is an attractor for a wide range of initial conditions. 

We close this section by noting that if the flux term is changed to $Q_a=\Gamma (u^b \phi_{,b})^n\phi_{,a}$,  for $n>1$, it can be shown that the updated equations of motion admit a similar attractor so long as $\Gamma\propto H^{2-n}$. This alleviates the finely-tuned time dependence of $\Gamma$ required for the stability of the background solution.

\section{Perturbations}\label{pertsec}
In this section, we study scalar perturbations to linear order about the background solution described in Eqs.~\eqref{bgsolutions}. We show that there exists a wide range in parameter space (\emph{i.e.,} choices of $c$ and $\gamma$) for which the comoving curvature perturbation acquires a scale invariant power spectrum. This result is displayed in Fig.~\ref{ns}; it is the main result of this paper.  

A full derivation of the scalar perturbation equations in spatially-flat gauge is presented in Appendix~\ref{pertderiv}.  In this gauge, all perturbed quantities can be expressed in terms of  the scalar potentials of the four-velocities of the radiation fluid, $\delta u_r$, and of the scalar field, $\delta u_\phi$.  In particular, the comoving curvature perturbation is
\begin{eqnarray}
\label{comocupe}
\mathcal{R}\equiv -\frac{H}{2\epsilon}(6x_0^2\delta u_\phi+4z_0^2\delta u_r).
\end{eqnarray} 
These potentials satisfy the coupled system
\begin{eqnarray}
\delta u_\phi''+\frac{C_1}{\tau}\delta u_\phi'+\left(k^2+\frac{C_2}{\tau^2}\right)\delta u_\phi &=&\mathcal{J}_r(\delta u_r,k,\tau)\nonumber\\
&&+\xi(k,\tau),\label{deltaconf2}\\
\delta u_r''+\frac{C_5}{\tau}\delta u_r'+\left(\frac{k^2}{3}+\frac{C_6}{\tau^2}\right)\delta u_r&=&\mathcal{J}_\phi(\delta u_\phi,k,\tau),\label{deltaurconf2}
\end{eqnarray}
where
\begin{eqnarray}
\mathcal{J}_r(\delta u_r,k,\tau)\equiv \frac{C_3}{\tau^2}\delta u_r+\frac{C_4}{\tau}\delta u_r',\label{jr}\\
\mathcal{J}_\phi(\delta u_\phi,k,\tau)\equiv \frac{C_7}{\tau^2}\delta u_\phi+\frac{C_8}{\tau}\delta u_\phi'\label{jphi},
\end{eqnarray}
 $\xi\equiv \Xi/\dot\phi^2$, and the $C_i$ are constants that depend on $c$ and $\gamma$, whose explicit definitions are given in Eqs.~\eqref{bigc} and \eqref{littlec1}-\eqref{littlec8}. 

The purpose of the rest of this section is to solve Eqs.~\eqref{deltaconf2} and \eqref{deltaurconf2} for general $c$ and $\gamma$ so that we can find $\mathcal{R}$ via Eq.~\eqref{comocupe}.  Before solving this system in general (Sec.~\ref{warmcase}), we first review the solution in the cold case when $\gamma=0$ (Sec.~\ref{coldcase}). 
\subsection{Cold Ekpyrosis}\label{coldcase}
In this subsection, we reproduce the result of standard, single-field ekpyrosis, \emph{i.e.,} without radiation, for which a scale-invariant spectrum for $\mathcal{R}$ is impossible \cite{Creminelli:2004jg}.

Recall that with no radiation ($\gamma=0$) and a sufficiently steep potential ($c>\sqrt{6}$), the background solution in Eqs.~\eqref{pinkx1} and \eqref{pinkz1} reduces to $(x_0,z_0)=(c/\sqrt{6},0)$.  As for the perturbations, $\delta u_r$ vanishes identically, and Eq.~\eqref{deltaconf2} becomes
\begin{eqnarray}
\label{sfeccp}
&&\delta u_\phi''+\frac{C_1}{\tau}\delta u_\phi'+\left(k^2 +\frac{C_2}{\tau^2}\right)\delta u_\phi=0,
\end{eqnarray}
with $C_1=-2$ and $C_2= 2 c^2 \left(c^2-3\right) \left(c^2-2\right)^{-2}$. The selection of Bunch-Davies vacuum fixes 
\begin{eqnarray}
\delta u_\phi(k,\tau)=\frac{\epsilon-1}{\sqrt{2\epsilon}}\sqrt{\frac{\pi}{4}}(-\tau)^{\frac{1-C_1}{2}}H_{\nu_\phi}^{(1)}(-k\tau),\label{sfesuper}
\end{eqnarray}
where, in terms of the function
\beq
\nu(X,Y)\equiv\frac{1}{2}\sqrt{(X-1)^2-4Y},
\eeq
we have defined $\nu_\phi\equiv \nu(C_1,C_2)$.
In the next subsection, we will use this same normalization for the perturbation, $\delta u_\phi$, since at early times, the temperature, $T$, and dissipation, $\Gamma$, are small. In the super-horizon limit, $-k\tau\to 0$, Eq.~\eqref{sfesuper} approaches $\delta u_\phi \propto k^{-\nu_\phi}$, so the spectral index is given by
\begin{eqnarray}
n_s&=&4-2\nu_\phi\nonumber\\
&=&3+\frac{4}{c^2-2}\label{coldtilt},
\end{eqnarray}
which is clearly blue (in particular $>3$) for ekpyrosis (which requires $c>\sqrt{6}$).  Thus, scale-invariance is impossible in the single-field model.  

\subsection{Warm Ekpyrosis}\label{warmcase}
In this subsection, we consider the ``warm'' case when $\gamma\neq 0$.  The presence of the radiation fluid introduces into Eqs.~\eqref{deltaconf2} and \eqref{deltaurconf2}  two crucial differences:  the first is that $C_1$ and $C_2$ depend not only on the steepness, $c$, of the potential, but also on the decay rate, $\gamma$; the second is that $\delta u_r$ is no longer negligible.

To solve the system, we decompose the scalar potential for the four-velocity of the radiation fluid as $\delta u_r=\delta u_r^{\rm h}+\delta u_r^{\rm p}$, where the first term is a homogeneous solution to Eq.~\eqref{deltaurconf2} with $\mathcal{J_\phi}$ set to $0$, \emph{i.e.,}
\beq
\delta u_r^{\rm h}(k,\tau) =(-k\tau)^{\frac{1-C_5}{2}}[a_1(k)J_{\nu_r}(-k\tau)+a_2(k)Y_{\nu_r}(-k\tau)],\label{durh}
\eeq
and the second term is the particular solution given by integrating over the retarded Green's function, \emph{i.e.,}
\beq
\delta u_r^{\rm p}(k,\tau)= k^{-1}\int_{-\infty}^\tau \,G_r(-k\tau,-k\bar{\tau})\mathcal{J}_\phi(\delta u_\phi,k,\bar{\tau}) d\bar{\tau}\label{gfs}.
\eeq
In the above, $a_1(k)$ and $a_2(k)$ are integration constants and 
\begin{eqnarray}
G_r(z,y)\equiv\frac{\pi}{2} y\left(z/y\right)^{\frac{1-C_5}{2}}&[J_{\nu_r}\left(z/\sqrt{3}\right) Y_{\nu_r}\left(y/\sqrt{3}\right)&\nonumber\\
-&Y_{\nu_r}\left(z/\sqrt{3}\right) J_{\nu_r}\left(y/\sqrt{3}\right)]&
\end{eqnarray}
with  $\nu_r\equiv \nu(C_5,C_6)$.
 In Appendix \ref{ap1}, we show that the integral in Eq.~\eqref{gfs} can be approximated by
\begin{equation}
\label{durp}
\delta u_r^{\rm p} \approx (C_7/C_6)\theta (1-k|\tau|) \delta u_\phi,
\end{equation}
where $\theta(x)$ is the Heaviside step function. That is, $\delta u_r^{\rm p}$ is negligible before horizon crossing and is a constant multiple of $\delta u_\phi$ after horizon crossing (see Fig.~\ref{pvz}). 

Armed with these solutions, we now turn to Eq.~\eqref{deltaconf2}.  The right side is a sum of three terms, $\mathcal{J}_r(\delta u_r^{\rm p},k,\tau)+\mathcal{J}_r(\delta u_r^{\rm h},k,\tau)+\xi$.  The last term is negligible as discussed in detail in Appendix~\ref{ap3}.  The second term is a rapidly decreasing function 
that depends on the initial state of $\delta u_r^{\rm h}$. We restrict attention to models where this term begins sufficiently small that it can be neglected.  Therefore, we need only consider $\mathcal{J}_r(\delta u_r^{\rm p})$.  Inside the horizon, it has no effect, but outside the horizon, it renormalizes the ``dissipation'' and ``frequency'' terms on the left side of Eq.~\eqref{deltaconf2} 
\begin{eqnarray}
C_1\to\tilde{C}_1&\equiv& C_1-C_4C_7C_6^{-1},\label{tildeC1}\\
C_2\to \tilde{C}_2&\equiv& C_2-C_3C_7C_6^{-1}\label{tildeC2},
\end{eqnarray}
as is clear from substituting Eq.~\eqref{durp} into Eq.~\eqref{jr} and putting the result into Eq.~\eqref{deltaconf2}.

\begin{figure}
\includegraphics[width=.45\textwidth]{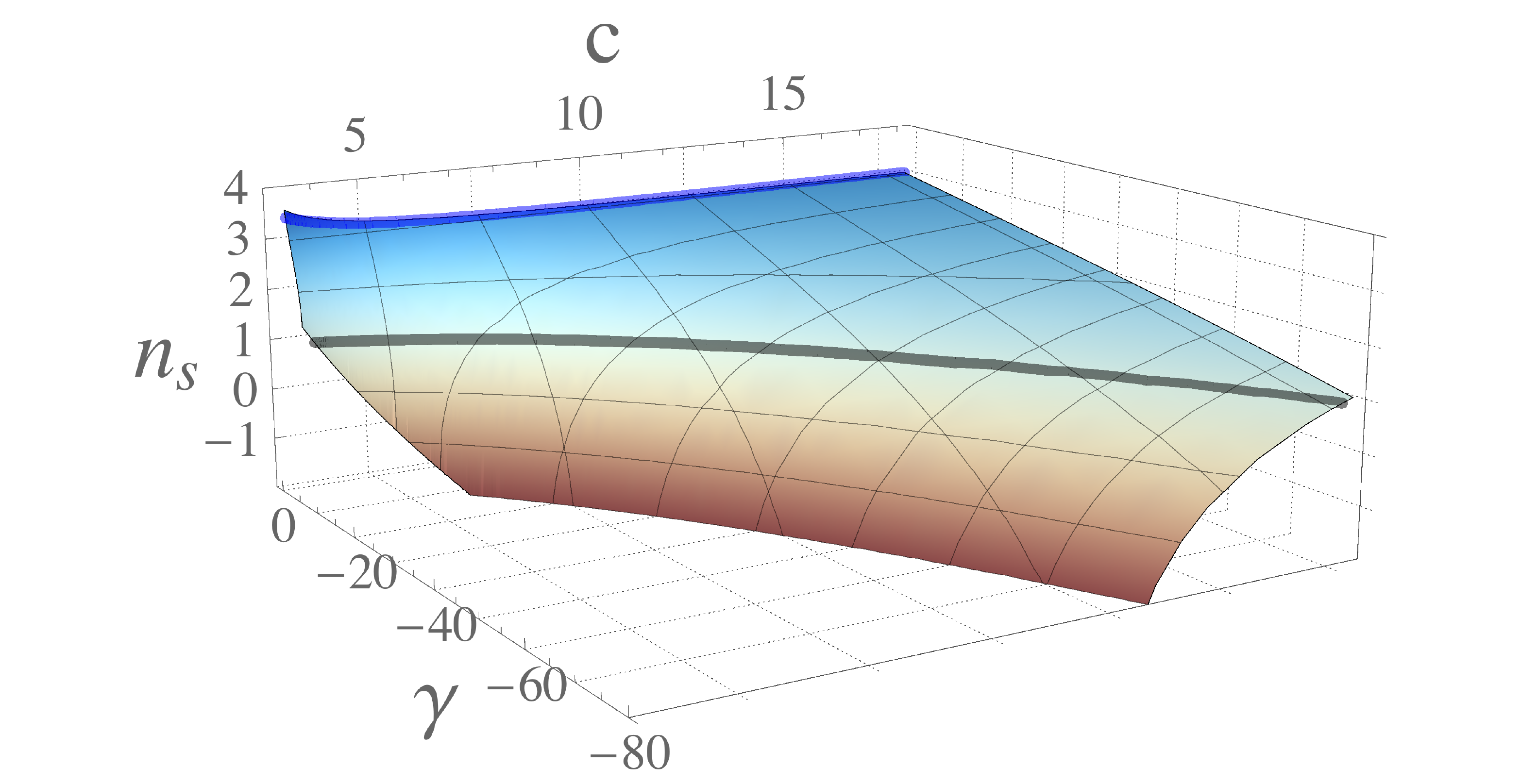}
\caption{\label{ns} This shows $n_s$ as a function of $c$ and $\gamma$ for the background solution in Eqs.~\eqref{bgsolutions}.}
\end{figure}

Therefore, the subhorizon solution is given by 
\begin{equation}
\label{subho}
\delta u_\phi^{\rm sub} = \frac{\epsilon-1}{\sqrt{2\epsilon}}\sqrt{\frac{\pi}{4}}(-\tau)^{\frac{1-C_1}{2}}H_{\nu_\phi}^{(1)}(-k\tau),
\end{equation} 
and the superhorizon solution is given by
\begin{equation}
\label{totsol}
\delta u_\phi^{\rm sup} = (-\tau)^{\frac{1-\tilde{C}_1}{2}}\left(\kappa_1 J_{\tilde\nu_\phi}(-k\tau)+\kappa_2 Y_{\tilde\nu_\phi}(-k\tau)\right),
\end{equation}
where $\tilde\nu_\phi\equiv \nu(\tilde{C}_1,\tilde{C}_2)$, and $\kappa_1$ and $\kappa_2$ are approximated by the following matching conditions at horizon crossing ($-k\tau=1)$
\begin{eqnarray}
\delta u_\phi^{\rm sub}&=&\delta u_\phi^{\rm sup},\\ 
(\delta u_\phi^{\rm sup})'-(\delta u_\phi^{\rm sub})' &=& -C_4C_7C_6^{-1} k\, \delta u_\phi^{\rm sub},
\end{eqnarray}
\emph{i.e.,}
\begin{eqnarray}
\kappa_1&=&-\frac{\pi ^{3/2} (\epsilon -1)}{8 \sqrt{2} C_6 \sqrt{\epsilon }} \bigg(H_{\nu_\phi }^{(1)}(1) (Y_{\tilde{\nu}_\phi}(1) (\tilde{C}_1-C_1\nonumber\\
&&+2(C_4 C_7C_6^{-1}+ \tilde{\nu}_\phi- \nu_\phi) )-2 Y_{\tilde{\nu}_\phi-1}(1))\nonumber\\
&&+2 H_{\nu_\phi -1}^{(1)}(1) Y_{\tilde{\nu}_\phi}(1)\bigg)\times k^{\frac{C_1-\tilde{C}_1}{2}}\label{kappa1},\\
\kappa_2&=&\frac{\pi ^{3/2} (\epsilon -1)}{8 \sqrt{2} C_6 \sqrt{\epsilon }} \bigg(H_{\nu_\phi }^{(1)}(1) (J_{\tilde{\nu}_\phi}(1) (\tilde{C}_1-C_1\nonumber\\
&&+2(C_4 C_7C_6^{-1}+ \tilde{\nu}_\phi- \nu_\phi) )-2 J_{\tilde{\nu}_\phi-1}(1))\nonumber\\
&&+2 H_{\nu_\phi -1}^{(1)}(1) J_{\tilde{\nu}_\phi}(1)\bigg)\times k^{\frac{C_1-\tilde{C}_1}{2}}\label{kappa2},
\end{eqnarray}
Substituting the solution in Eq.~\eqref{totsol}, together with Eqs.~\eqref{kappa1} and \eqref{kappa2}, into Eq.~\eqref{comocupe}, we find that the primordial power spectrum of the comoving curvature perturbation on superhorizon scales is given by
\begin{equation}
\label{ccp}
\Delta^2_\mathcal{R}(k,\tau)\equiv \frac{k^3}{2\pi^2}|\mathcal{R}|^2\approx \mathcal{O}(10^{-4})V_\text{end}^{\frac{1+\tilde{C}_1+2 \tilde\nu_\phi}{2}}k^{n_s-1},
\end{equation}
where $V_\text{end}\equiv |V_0|e^{-c\phi_{\rm e}}$ is the magnitude of the potential energy density when ekpyrosis ends, and the spectral index is given by
\begin{equation}
\label{warmtilt}
n_s=4-2\tilde{\nu}_\phi+(C_1-\tilde{C}_1),
\end{equation}
which is plotted in Fig.~\ref{ns}.  

Given any point in the $c$-$\gamma$ plane, the height of the surface above that point shows the spectral index, $n_s$. The color scheme reflects that for $n_s>1$, the spectrum is blue and for $n_s<1$, the spectrum is red. The thick, blue curve at $\gamma=0$ reproduces the results of ordinary, single-field ekpyrosis from Eq.~\eqref{coldtilt}.  As discussed in Sec.~\ref{coldcase}, this curve describes a blue-tilted spectrum that is inconsistent with observation. However, note the effect of particle production on the spectral index:   at any value of $c$, increasing the decay rate, $|\gamma|$, reddens the spectrum. In particular, the thick, black curve has $n_s=1$. Any choice of $c$ and $\gamma$ along this curve corresponds to an exactly scale-invariant spectrum. For such a choice, the exponent of $V_\text{end}$ in Eq.~\eqref{ccp} can be computed and is roughly $.61$, so that to match the observed amplitude, $V_\text{end}$ must be made of order $V_\text{end}^{1/4}\sim 10^{16}$ GeV, which is high enough to recover the successful predictions of hot big bang nucleosynthesis.

\section{Discussion}\label{conc}
In this work, we have presented a scenario for ekpyrosis that continuously generates a scale-invariant spectrum of adiabatic perturbations. The key is the continuous decay of the ekpyrotic field; this decay introduces a friction term that allows a scale-invariant spectrum to be achieved. More generally, as can be seen by following a curve of constant $c$ along the surface in Fig.~\ref{ns}, we showed that the effect of particle production is to redden the power spectrum of the supercooled theory.  

We view the elimination of the second scalar field and hence any subsequent conversion mechanism as a major simplification, and a return to the spirit of the original formulation of ekpyrosis, since the hydrodynamical behavior at finite temperature is universal regardless of the details of its microscopic origin. While we have not attempted to embed this phase into a complete cosmological history, the decay into radiation presents the tantalizing possibility of evading the need for additional reheating. 

There are two key assumptions that merit attention. The first is that the decay rate must scale with the Hubble parameter. This scaling represents the greatest source of fine-tuning (although see the last paragraph of Sec.~\ref{backgroundsec} for a possible alternative). The second is that the initial fluctuations of the fluid, $\delta u_r^{\rm h}$, are small enough to be neglected. If these conditions are met, it is always possible to choose the parameters $c$ and $\gamma$ such that a scale invariant spectrum is achieved.

There are many directions for future work. One possibility is to consider generalizations of the radiation fluid within the framework presented here, as was done in warm inflation \cite{Bastero-Gil:2014jsa}. For example, one could analyze a fluid with a non-relativistic equation of state or that is out of thermal equilibrium.  One could also include viscosity by adding corrections to its energy-momentum tensor, $T^{(r)}_{ab}$.  Another possibility is to devise a microphysical theory-- to identify the microscopic degrees of freedom comprising the fluid that realizes the effective dynamics described here. For example, one could try to reproduce the trapped inflation scenario in a contracting universe \cite{Green:2009ds}.  In warm inflation, this is difficult, though not impossible \cite{ Berera:1998px, Berera:1998gx,Bartrum:2013fia}, to achieve because, as argued in Ref.~\cite{Yokoyama:1998ju}, the dissipation coefficient, $\Gamma$, appears as the result of a small correction to a sub-leading thermal correction to the potential energy density of the inflaton, which must be extremely flat to support inflation. Thus, non-negligible $\Gamma$ requires large thermal corrections, which spoil the extreme flatness of the potential. In ekpyrosis, such extreme flatness is neither required nor permitted.

It is also important to compute the non-Gaussian signatures for this model to verify that they are small. A reason to be optimistic is that, as we have shown, neither the steepness, $c$, of the potential nor the equation of state parameter of the universe, $\epsilon$, needs to be tuned particularly large. For comparison, the models discussed in Refs.~\cite{2013PhLB..724..192L,Li:2014yla,Ijjas:2014fja,Levy:2015awa} generate no non-Gaussianity during the ekpyrotic phase. 

We thank L. Berezhiani, R. Brandenberger, A. Matas, D. Spergel, P. Steinhardt, and N. Turok for useful discussions. We especially thank P. Steinhardt for useful suggestions to improve the manuscript.

\begin{appendix}
\section{Perturbation Equations}\label{pertderiv}
In this appendix, we will derive the linearized perturbation equations. We will follow the notation of Ref.~\cite{Hwang:2001fb} (see also Ref.~\cite{Malik:2004tf}). To simplify the derivation, we will find the linearized equations for the ensemble expectation values of the fields. This implies that any stochastic contribution to these equations will vanish. 

A metric with the most general scalar-type perturbation in a flat Friedman-Robertson-Walker background is
\begin{eqnarray}
ds^2=&&-a^2(1+2\alpha)d\tau^2-2a^2\beta_{,i}d\tau dx^i\nonumber\\
&&+a^2\left[\delta_{ij}(1+\varphi)+2\psi_{,ij}\right]dx^idx^j
\end{eqnarray}
Ignoring anisotropic stress, the energy-momentum tensor for the fluid can be decomposed as
\begin{eqnarray}
T^{(r)\tau}_{\,\,\,\,\,\,\,\,\,\,\,\tau}&=&-(\rho_r+\delta\rho_r)\\
T^{(r)\tau}_{\,\,\,\,\,\,\,\,\,\,\,i}&=&a (\rho_r+p_r)\delta u_{r,i}\\
T^{(r)i}_{\,\,\,\,\,\,\,\,\,\,\,j}&=&(p_r+\delta p_r)\delta_{ij},
\end{eqnarray}
Thus, perturbations in the fluid are parameterized by $\delta \rho_r,\delta p_r$ and $\delta u_r$.  For simplicity, we will assume $\delta p_r =\delta\rho_r/3$, though this is not central to our results.  In writing the perturbation equations, it is useful to define the shear, $\chi\equiv a(\beta+a \dot{\psi})$, and the perturbed expansion of the normal-frame vector field  $\kappa \equiv 3(-\dot{\varphi}+H\alpha)+\frac{k^2}{a^2}\chi$. In Fourier space, the perturbation equations are 
{\small
\begin{eqnarray}
-\frac{k^2}{a^2}\varphi+H\kappa&=&-\frac{1}{2}\delta\rho \label{hw11},\\
\kappa-\frac{k^2}{a^2}\chi+\frac{3}{2}\sum_{i=r,\phi}(\rho_i+p_i)\delta u_i&=&0 \label{hw12},\\
\dot\chi+H\chi-\alpha-\varphi&=&0 \label{hw13}, \\
\dot{\kappa}+2H\kappa+\left(3\dot{H}-\frac{k^2}{a^2}\right)\alpha&=&\frac{1}{2}(\delta\rho+3\delta p)\label{hw14},\\
\delta\dot\rho_r+3H(\delta\rho_r+\delta p_r)&=&-\frac{k^2}{a^2}(\rho_r+p_r)\delta u_r+\delta q_{r}\nonumber\\
&&+\dot{\rho}_r\alpha+(\rho_r+p_r)\kappa \label{hw15},\\
\frac{-1}{a^3(\rho_r+p_r)}\frac{d}{dt}\left[a^3(\rho_r+p_r)\delta u_r\right]&=&\frac{\delta p_r}{\rho_r+p_r}+\alpha\nonumber\\
&&-\frac{j_r}{\rho_r+p_r},\label{hw16}\\
\delta\ddot\phi+3H\delta\dot\phi+\left(\frac{k^2}{a^2}+V_{,\phi\phi}\right)\delta\phi&=&\dot\phi(\kappa+\dot\alpha)-\delta q_\phi\nonumber\\
&&+(2\ddot\phi+3H\dot\phi)\alpha \label{hw102},
\end{eqnarray}}
where
\begin{eqnarray}
\delta\rho&\equiv& \delta\rho_r +\dot{\phi}\delta\dot\phi-\dot\phi^2\alpha+V_{,\phi}\delta\phi,\\
\delta p &\equiv&\delta p_r+\dot\phi\delta\dot\phi-\dot\phi^2\alpha-V_{,\phi}\delta\phi,\\
\delta u_\phi&\equiv&-\delta\phi/\dot\phi,\\
\delta q_{r}&\equiv&\delta\Gamma\dot\phi^2+2\Gamma\dot\phi\delta\dot\phi-2\alpha\Gamma\dot\phi^2,\\
\delta q_{\phi}&\equiv&\delta \Gamma \dot\phi-\Gamma\alpha\dot\phi+\Gamma\delta\dot\phi,\\
j_r&\equiv&-\Gamma\dot\phi\delta\phi.
\end{eqnarray}
Eqs.~\eqref{hw11}-\eqref{hw16} are, respectively, the $G^t_t$ component of the field equations, the $G^t_i$ component, the $G^i_j-\frac{1}{3}\delta^i_j G^k_k$ component, the $G^i_i-G^t_t$ component, the $T^{(r)b}_{\,\,\,\,\,\,\,\,\,\,\,i;b}=Q_i$ component, the $T^{(r)b}_{\,\,\,\,\,\,\,\,\,\,\,t;b}=Q_t$ component, and the $T^{(\phi)b}_{\,\,\,\,\,\,\,\,\,\,\,t;b}=-Q_t$ component. 

Henceforth, we work in spatially flat gauge ($G=\varphi=0$).  Then Eqs.~\eqref{hw11} and \eqref{hw12} can be solved algebraically for the metric variables $\alpha$ and $\beta$ in terms of the matter variables $\delta\phi,\delta u_r,$ and $\delta\rho_r$. Eq.~\eqref{hw16} can then be solved algebraically for $\delta \rho_r$ in terms of $\delta u_r$ and $\delta\phi$.  Substituting these results into Eqs.~\eqref{hw15} and \eqref{hw102} leaves two closed equations for the variables $\delta u_r$ and $\delta u_\phi$.  Specializing to the background solution in Eqs.~\eqref{bgsolutions}, these are
\begin{eqnarray}
\delta \ddot{u}_\phi&+&c_1H\delta \dot{u}_\phi+\left(\frac{k^2}{a^2}+c_2H^2\right)\delta u_\phi\nonumber\\
&=&c_3H^2\delta u_r+c_4H\delta \dot{u}_r \label{delta}\\
\delta \ddot{u}_r&+&c_5 H\delta \dot {u}_r+\left(\frac{k^2}{3a^2}+c_6H^2\right)\delta u_r\nonumber\\
&=&c_7 H^2\delta u_\phi +c_8 H \delta \dot{u}_\phi \label{deltaur},
\end{eqnarray}
with the constants $c_i$ defined by
\begin{eqnarray}
c_1&\equiv&-\frac{\sqrt{6} c z_0^2}{x_0}-\sqrt{6} c x_0+\frac{\sqrt{6} c}{x_0}-\gamma -3,\label{littlec1}\\
c_2&\equiv&6 \sqrt{6} c x_0^3+6 \sqrt{6} c x_0 z_0^2-6 \sqrt{6} c x_0-18 x_0^4\nonumber\\
&&-24 x_0^2 z_0^2+27 x_0^2+2 \gamma  z_0^2+6 z_0^2,\label{littlec2}\\
c_3&\equiv&-\frac{2 \sqrt{6} c z_0^4}{x_0}-2 \sqrt{6} c x_0 z_0^2+\frac{2 \sqrt{6} c z_0^2}{x_0}\nonumber\\
&&-8 \gamma  x_0^2+12 x_0^2 z_0^2+16 z_0^4-2 \gamma  z_0^2-8 z_0^2,\label{littlec3}\\
c_4&\equiv&-4 z_0^2, \label{littlec4}\\
c_5&\equiv&\frac{4 \gamma  x_0^2}{z_0^2}-1, \label{littlec5}\\
c_6&\equiv&-\frac{2 \sqrt{6} c \gamma  x_0^3}{z_0^2}-2 \sqrt{6} c \gamma  x_0+\frac{2 \sqrt{6} c \gamma  x_0}{z_0^2}\nonumber\\
&&-\frac{6 \gamma  x_0^4}{z_0^2}-7 \gamma  x_0^2-\frac{4 \gamma ^2 x_0^2}{z_0^2}-\frac{6 \gamma  x_0^2}{z_0^2}\nonumber\\
&&-8 x_0^2 z_0^2+3 x_0^2-8 z_0^4+10 z_0^2, \label{littlec6}\\
c_7&\equiv&-\frac{2 \sqrt{6} c \gamma  x_0^3}{z_0^2}-3 \sqrt{6} c x_0^3-2 \sqrt{6} c \gamma  x_0+\frac{2 \sqrt{6} c \gamma  x_0}{z_0^2}\nonumber\\
&&-3 \sqrt{6} c x_0 z_0^2+3 \sqrt{6} c x_0+12 x_0^4-5 \gamma  x_0^2-\frac{4 \gamma ^2 x_0^2}{z_0^2}\nonumber\\
&&-\frac{6 \gamma  x_0^2}{z_0^2}+12 x_0^2 z_0^2-18 x_0^2, \label{littlec7}\\
c_8&\equiv&\frac{5 \gamma  x_0^2}{2 z_0^2}+2 x_0^2 \label{littlec8},.
\end{eqnarray}
For concreteness, we have assumed $\Gamma \propto \sqrt{V(\phi)}$, independent of $\rho_r$ and $\dot\phi$, \emph{i.e.,} $\Gamma=-\gamma \sqrt{-V(\phi)/(3(x_0^2+z_0^2-1))}.$ For this choice, $\delta\Gamma=\gamma cH\delta\phi/2$. 


As we explained above, in deriving these equations, we have averaged out the stochastic fluctuations. However, at background level, we know that whatever microphysical process generates the dissipation, $\Gamma$, in the equation of motion for the scalar field, must also be accompanied by a stochastic source, $\Xi$, whose correlation satisfies the Fluctuation-Dissipation Theorem (see Appendix~\ref{ap3} for details). That is, Eq.~\eqref{delta} must be replaced with
\bea
\label{delta2}
\delta \ddot{u}_\phi&+&c_1H\delta \dot{u}_\phi+\left(\frac{k^2}{a^2}+c_2H^2\right)\delta u_\phi \nonumber \\
&=&c_3H^2\delta u_r+c_4H\delta \dot{u}_r+\xi(k,t),
\ea
where $\xi\equiv \Xi/\dot{\phi}^2$ with the extra factors of $\dot\phi$ in the denominator coming from the change of variables from $\delta\phi$ to $\delta u_\phi$. In conformal time, Eqs.~\eqref{delta2} and \eqref{deltaur} become Eqs.~\eqref{deltaconf2} and \eqref{deltaurconf2} ,
with 
\begin{equation}
\label{bigc}
 C_i =
  \begin{cases} 
      \hfill \frac{c_i-1}{\epsilon-1}    \hfill & \text{ if $i=1,5$} \\
      \hfill \frac{c_i}{(\epsilon-1)^2} \hfill & \text{ if $i=2,3,6,7$} \\
      \hfill \frac{c_i}{\epsilon-1} \hfill & \text{ if $i=4,8$}
  \end{cases}
\end{equation}
where again $\epsilon\equiv 3x_0^2+2z_0^2$.
\section{Locality of fluid response}\label{ap1}
Since, in general, fluids behave non-locally we will show in this appendix how locality can be recovered in the late time limit. This is because the sources for the fluid are $\tau^{-2} \delta u_\phi(k,\tau) , \tau^{-1} \delta u'_\phi(k,\tau)$ and not $\delta u_\phi$ itself. Using the expression for the source, $\mathcal{J}_\phi$, in Eq.~\eqref{jphi} and integrating the derivative term by parts, the particular solution for the radiation fluid in Eq.~\eqref{gfs} can be rewritten as 
\beq
\label{urpsol}
\delta u_r^{\rm p}(z) = \int_z^\infty dy K_r(z,y) \delta u_\phi (y),
\eeq
where $z\equiv-k\tau$ and we defined the kernel
\beq
K_r(z,y) \equiv\frac{C_7+C_8}{y^2} G_r(z,y) - \frac{C_8}{y} G_{r,y}(z,y).
\eeq
\begin{figure}
\subfigure[]{
\includegraphics[width=.45\textwidth]{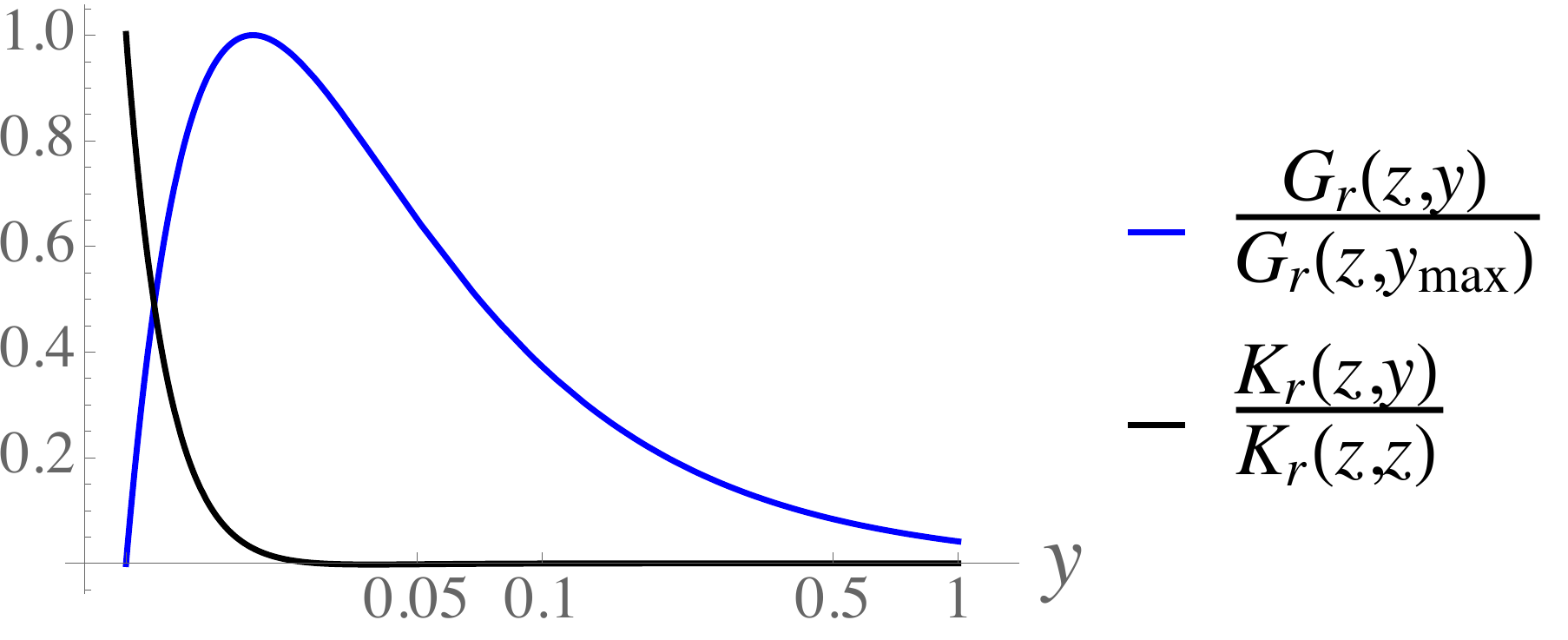}}
\subfigure[]{
\includegraphics[width=.45\textwidth]{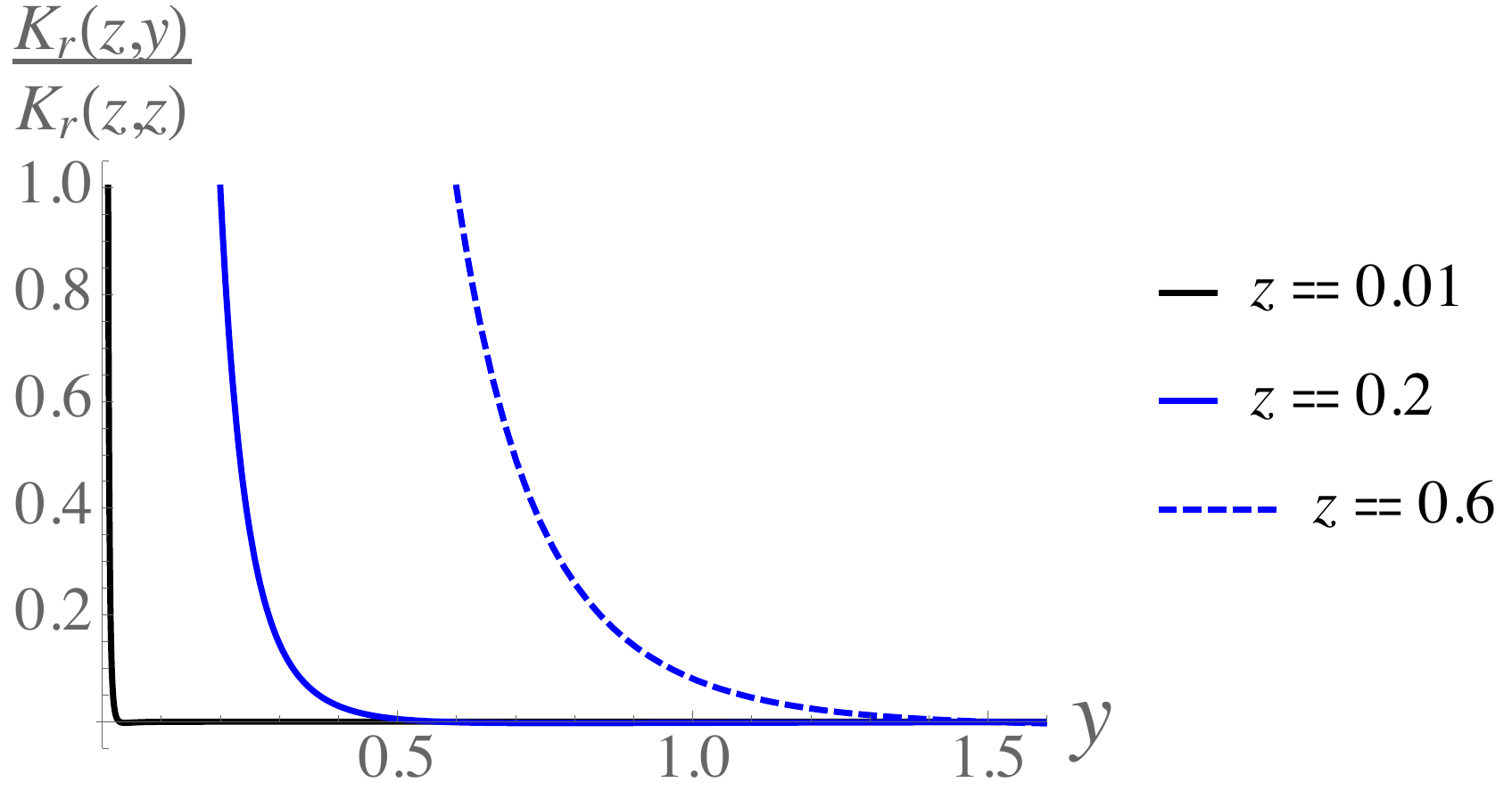}}
\caption{\label{locality}(a) Comparison of the locality of the Green's function, $G_r$, (blue) and the kernel, $K_r$, (black) for $z=10^{-2}$. Both are normalized such that their maximum value is $1$:  as should be clear, $y_\text{max}$ is the argument for which $G_r(z,y)$  is maximized. Note the logarithmic scale on the horizontal axis. \\(b) Comparison of the fluid kernel, $K_r$ at different final times, $z$. As modes are stretched beyond the horizon $z\ll 1$, this kernel becomes increasingly local.  For both plots, we used $(c,\gamma)=(15,-56.8)$.}
\end{figure}

\begin{figure}
\includegraphics[width=.45\textwidth]{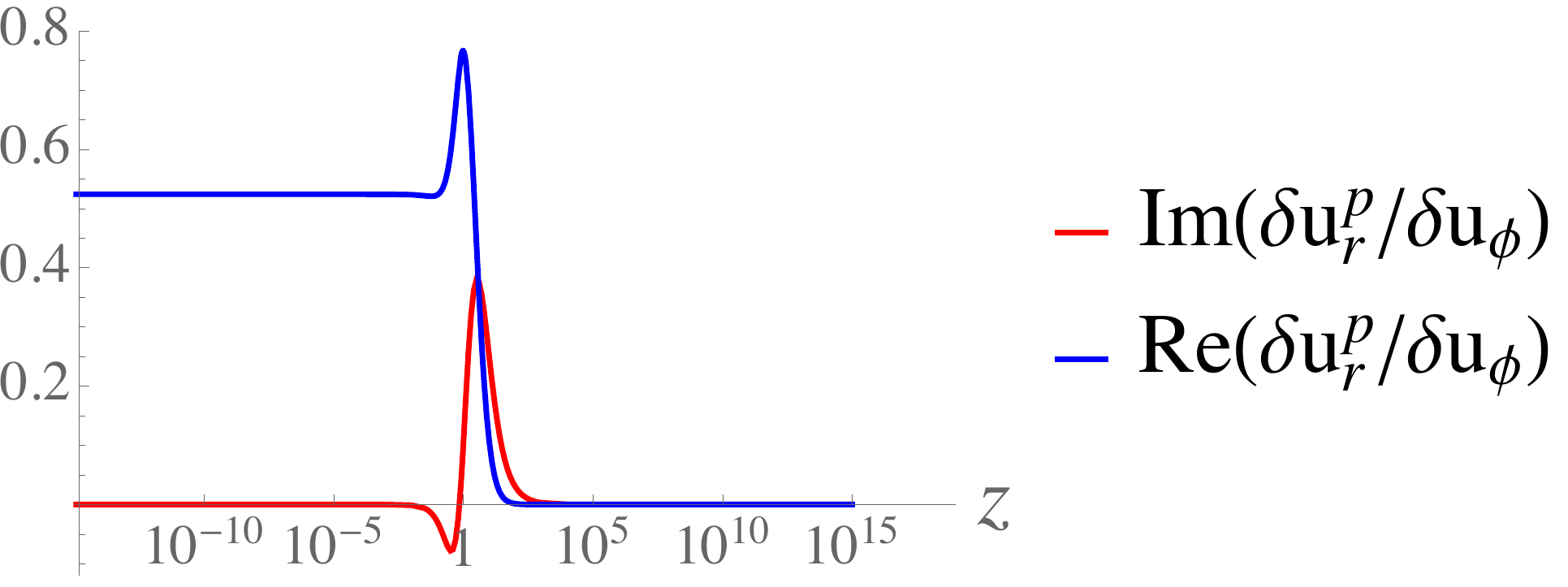}
\caption{\label{pvz} This plot shows the time dependence of the real (blue) and imaginary (red) parts of the ratio $\delta u_r^{\rm p}/\delta u_\phi$ when $(c,\gamma)=(15,-56.8)$.  Recall $\delta u_\phi$ is given by Eqs.~\eqref{subho} and \eqref{totsol} and $\delta u_r^{\rm p}$ is given by Eq.~\eqref{gfs}, or, equivalently, by Eq.~\eqref{urpsol}. There is a sharp transition once a mode exits the horizon.  Inside the horizon ($z>1$), the particular solution for the fluid $\delta u_r^{\rm p}$ is negligible.  Outside ($z<1$) it rapidly approaches a constant factor, roughly $C_7/C_6$, times $\delta u_\phi$.  This justifies the mode matching procedure in the text.}
\end{figure}

Now, we will show that this kernel behaves locally in the small $z$ (superhorizon) approximation. It follows from the explicit expression of the Green's functions that $K_r(z,z)\rightarrow \infty$ and $K_r(z,y)\rightarrow 0$ for $z\neq y$ as $z\rightarrow 0$. These properties are illustrated in Fig.~\ref{locality}. To compute the particular solution for the radiation fluid, we can therefore make the local approximation 
\beq
\label{urpkern}
\delta u_r^{\rm p}(z) = \int_z^\infty dy K_r(z,y) \delta u_\phi (y) \approx \delta u_\phi(z) \int_z^\infty dy K_r(z,y).
\eeq
In the small $z$ limit, this integral can be done exactly and gives
\beq
 \int_z^\infty dy K_r(z,y) = \frac{C_7}{C_6} + \calo(z^{\frac{1-C_5+2\nu_r}{2}}).
\eeq 
Therefore, in this limit, we make the approximation (see Fig.~\ref{pvz})
\beq
\delta u_r^{\rm p}(z) \approx (C_7/C_6)\delta u_\phi(z) + \calo(z^{\frac{1-C_5+2\nu_r}{2}}).
\eeq

\section{Thermal contribution to the scalar spectrum}\label{ap3}
In this appendix, we will show that the thermal contribution to the power spectrum of the comoving curvature perturbation, $\Delta_\mathcal{R}^2$, is negligible for the observable modes in comparison to the vacuum, scale-invariant contribution. 

To prove this, we write the particular solution for the scalar field perturbation in terms of the stochastic noise 
\beq
\label{noiseu}
\delta u_\phi (\mathbf{k},\tau_{\rm e})= k^{-1} \int_{-\infty}^{\tau_{\rm e}} d\tau G_\phi (-k\tau_{\rm e},-k\tau) \xi(\mathbf{k},\tau),
\eeq
where $G_\phi$ is the retarded Green's function for Eq.~\eqref{deltaconf2}, and again, $\tau_{\rm e}$ is the time at which ekpyrosis ends. The two-point function of the noise follows from the Fluctuation-Dissipation Theorem in Eq.~\eqref{noiseterms} 
\beq
\lb \xi(\textbf{k},\tau) \xi(\textbf{k}',\tau')\rb = \mathcal{N}_{\rm FD}(2\pi)^3 \delta(\textbf{k}+\textbf{k}')\delta (\tau-\tau'),\label{fdback}
\eeq
where the noise kernel is given by 
\beq 
\mathcal{N}_{\rm FD} \equiv 2\Gamma T/\dot{\phi}^2,
\eeq
with the extra factors of $\dot{\phi}$ in the denominator coming from the change of variables from $\delta \phi$ to $\delta u_\phi$. Substituting Eq.~\eqref{fdback}  into Eq.~\eqref{noiseu} and changing the integration variable to $y=-k\tau$, the thermal power spectrum for $\delta u_{\phi}$ is given by 
\begin{eqnarray}
&&\lb \delta u_\phi (\mathbf{k},\tau_{\rm e})\delta u_\phi (\mathbf{k}',\tau_{\rm e})\rb \nonumber\\
&&= \frac{1}{k^3} \left\{\int_{-k\tau_{\rm e}}^\infty [G_\phi(-k\tau_{\rm e},y)]^2 \mathcal{N}_{\rm FD} dy \right\}\times\nonumber\\
&&(2\pi)^3 \delta^{(3)}(\mathbf{k}+\mathbf{k}'). \label{tpf12}
\end{eqnarray}
To compute this integral, we separate out the time dependence of the noise kernel, \emph{i.e.,} $\mathcal{N}_{\rm FD}=\mathcal{N}_0 (-\tau)^{\frac{1}{2}+\frac{1}{2(\epsilon-1)}}$ with 
\beq
\mathcal{N}_{0}\equiv\frac{-\gamma(\epsilon-1)^{1/2}}{3x_0^2}\left(\frac{45}{\pi^2}z_0^2\right)^{1/4}(-\tau_{\rm e})^{-\frac{1}{2(\epsilon-1)}}.\label{nb0}
\eeq
For the Green's function, $G_\phi$, we use the same approximation we used to compute $\delta u_\phi$ in Eqs.~\eqref{subho} and \eqref{totsol}, namely finding the solutions for $y>1$ and $y<1$ and matching at horizon crossing, taking account of the effect of $\mathcal{J}_r$ according to the change in dissipation and frequency in Eqs.~\eqref{tildeC1} and \eqref{tildeC2}. 

\begin{figure}[h!]
\includegraphics[width=.45\textwidth]{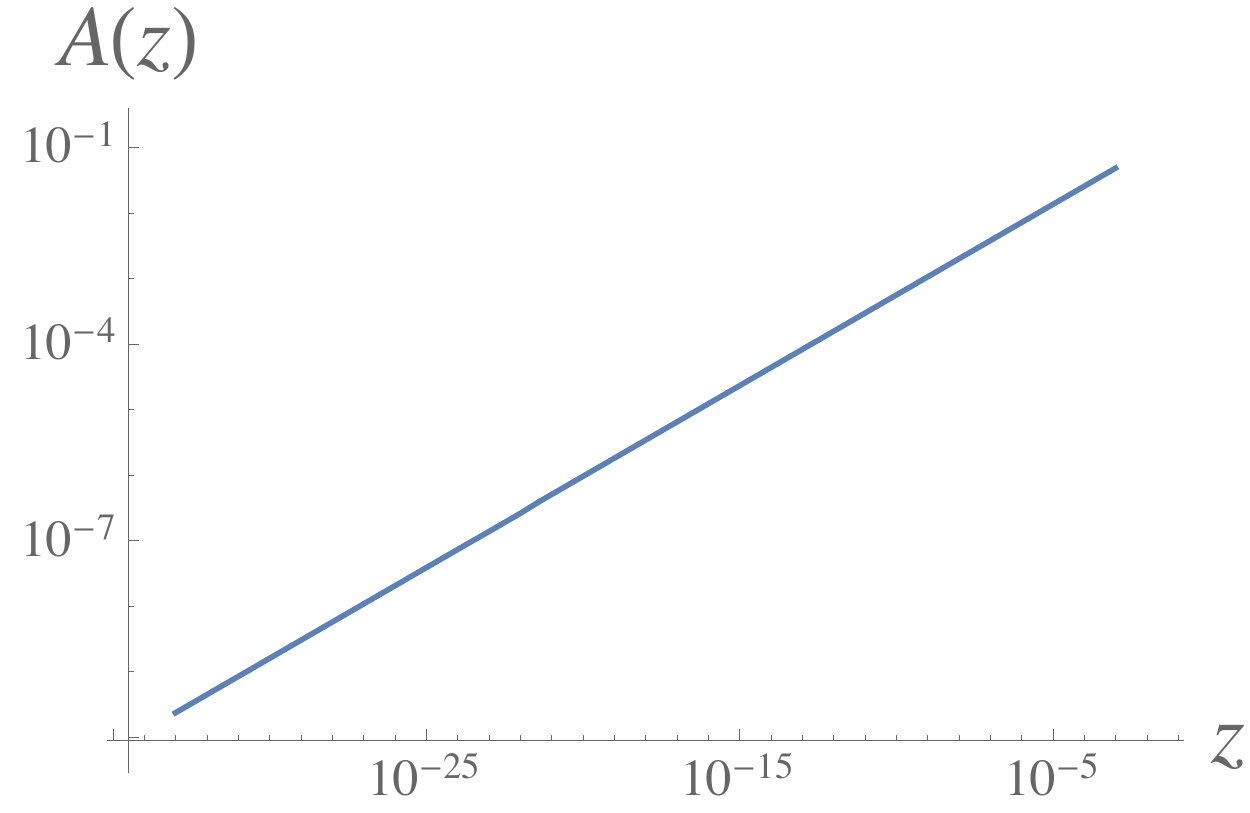}
\caption{\label{bt2p} This plot shows the result for $\mathcal{A}(z)$ obtained by matching the different solutions at horizon crossing. The largest scales correspond to the smallest values of $z$. Thus, there is a strong suppression of the thermal contribution to $\Delta_\mathcal{R}^2$ on the largest scales.  Again, we have used the parameter choice $(c,\gamma)=(15,-56.8)$.}
\end{figure}

Then, dropping the arguments on the left side of Eq.~\eqref{tpf12}, the power spectrum is
\begin{eqnarray}
k^3 \lb \delta u_\phi^2\rb &=&\mathcal{N}_0 (-\tau_{\rm e})^{\frac{1}{2}+\frac{1}{2(\epsilon-1)}} \mathcal{A}(-k\tau_{\rm e})\times\nonumber\\
&&(2\pi)^3 \delta^{(3)} (\textbf{k}+\textbf{k}') \label{pb}
\end{eqnarray}
where the function
\beq
\mathcal{A}(z)\equiv z^{-\frac{1}{2}-\frac{1}{2(\epsilon-1)}}\int_{z}^\infty dy[G_\phi(z,y)]^2  y^{\frac{1}{2}+\frac{1}{2(\epsilon-1)}}
\eeq 
is plotted in Fig.~\ref{bt2p}.  Using Eqs.~\eqref{nb0}, \eqref{pb}, and \eqref{comocupe} in Eq.~\eqref{ccp}, we find that this contribution to $\Delta^2_\mathcal{R}$ is suppressed relative to the vacuum result by a factor that is weakly dependent on $c$ and $\gamma$ and is of order $\mathcal{A}(-k\tau_{\rm e})V_{\text{end}}^{.14}$, \emph{e.g.}, for the choice $(c,\gamma)=(15,-56.8)$. Thus, the ratio of this thermal contribution to the vacuum contribution is of order $\mathcal{O}(10^{-9})$ for the largest observable modes.

\end{appendix}

\bibliography{bibliography}

\begin{thebibliography}{53}%
\makeatletter
\providecommand \@ifxundefined [1]{%
 \@ifx{#1\undefined}
}%
\providecommand \@ifnum [1]{%
 \ifnum #1\expandafter \@firstoftwo
 \else \expandafter \@secondoftwo
 \fi
}%
\providecommand \@ifx [1]{%
 \ifx #1\expandafter \@firstoftwo
 \else \expandafter \@secondoftwo
 \fi
}%
\providecommand \natexlab [1]{#1}%
\providecommand \enquote  [1]{``#1''}%
\providecommand \bibnamefont  [1]{#1}%
\providecommand \bibfnamefont [1]{#1}%
\providecommand \citenamefont [1]{#1}%
\providecommand \href@noop [0]{\@secondoftwo}%
\providecommand \href [0]{\begingroup \@sanitize@url \@href}%
\providecommand \@href[1]{\@@startlink{#1}\@@href}%
\providecommand \@@href[1]{\endgroup#1\@@endlink}%
\providecommand \@sanitize@url [0]{\catcode `\\12\catcode `\$12\catcode
  `\&12\catcode `\#12\catcode `\^12\catcode `\_12\catcode `\%12\relax}%
\providecommand \@@startlink[1]{}%
\providecommand \@@endlink[0]{}%
\providecommand \url  [0]{\begingroup\@sanitize@url \@url }%
\providecommand \@url [1]{\endgroup\@href {#1}{\urlprefix }}%
\providecommand \urlprefix  [0]{URL }%
\providecommand \Eprint [0]{\href }%
\providecommand \doibase [0]{http://dx.doi.org/}%
\providecommand \selectlanguage [0]{\@gobble}%
\providecommand \bibinfo  [0]{\@secondoftwo}%
\providecommand \bibfield  [0]{\@secondoftwo}%
\providecommand \translation [1]{[#1]}%
\providecommand \BibitemOpen [0]{}%
\providecommand \bibitemStop [0]{}%
\providecommand \bibitemNoStop [0]{.\EOS\space}%
\providecommand \EOS [0]{\spacefactor3000\relax}%
\providecommand \BibitemShut  [1]{\csname bibitem#1\endcsname}%
\let\auto@bib@innerbib\@empty
\bibitem [{\citenamefont {Guth}(1981)}]{PhysRevD.23.347}%
  \BibitemOpen
  \bibfield  {author} {\bibinfo {author} {\bibfnamefont {A.~H.}\ \bibnamefont
  {Guth}},\ }\href {\doibase 10.1103/PhysRevD.23.347} {\bibfield  {journal}
  {\bibinfo  {journal} {Phys. Rev. D}\ }\textbf {\bibinfo {volume} {23}},\
  \bibinfo {pages} {347} (\bibinfo {year} {1981})}\BibitemShut {NoStop}%
\bibitem [{\citenamefont {Albrecht}\ and\ \citenamefont
  {Steinhardt}(1982)}]{Albrecht:1982wi}%
  \BibitemOpen
  \bibfield  {author} {\bibinfo {author} {\bibfnamefont {A.}~\bibnamefont
  {Albrecht}}\ and\ \bibinfo {author} {\bibfnamefont {P.~J.}\ \bibnamefont
  {Steinhardt}},\ }\href {\doibase 10.1103/PhysRevLett.48.1220} {\bibfield
  {journal} {\bibinfo  {journal} {Phys.Rev.Lett.}\ }\textbf {\bibinfo {volume}
  {48}},\ \bibinfo {pages} {1220} (\bibinfo {year} {1982})}\BibitemShut
  {NoStop}%
\bibitem [{\citenamefont {Linde}(1982)}]{Linde:1981mu}%
  \BibitemOpen
  \bibfield  {author} {\bibinfo {author} {\bibfnamefont {A.~D.}\ \bibnamefont
  {Linde}},\ }\href {\doibase 10.1016/0370-2693(82)91219-9} {\bibfield
  {journal} {\bibinfo  {journal} {Phys.Lett.}\ }\textbf {\bibinfo {volume}
  {B108}},\ \bibinfo {pages} {389} (\bibinfo {year} {1982})}\BibitemShut
  {NoStop}%
\bibitem [{\citenamefont {Khoury}\ \emph {et~al.}(2002)\citenamefont {Khoury},
  \citenamefont {Ovrut}, \citenamefont {Seiberg}, \citenamefont {Steinhardt},\
  and\ \citenamefont {Turok}}]{Khoury:2001bz}%
  \BibitemOpen
  \bibfield  {author} {\bibinfo {author} {\bibfnamefont {J.}~\bibnamefont
  {Khoury}}, \bibinfo {author} {\bibfnamefont {B.~A.}\ \bibnamefont {Ovrut}},
  \bibinfo {author} {\bibfnamefont {N.}~\bibnamefont {Seiberg}}, \bibinfo
  {author} {\bibfnamefont {P.~J.}\ \bibnamefont {Steinhardt}}, \ and\ \bibinfo
  {author} {\bibfnamefont {N.}~\bibnamefont {Turok}},\ }\href {\doibase
  10.1103/PhysRevD.65.086007} {\bibfield  {journal} {\bibinfo  {journal}
  {Phys.Rev.}\ }\textbf {\bibinfo {volume} {D65}},\ \bibinfo {pages} {086007}
  (\bibinfo {year} {2002})},\ \Eprint {http://arxiv.org/abs/hep-th/0108187}
  {arXiv:hep-th/0108187 [hep-th]} \BibitemShut {NoStop}%
\bibitem [{\citenamefont {Komatsu}\ \emph {et~al.}(2009)\citenamefont {Komatsu}
  \emph {et~al.}}]{Komatsu:2008hk}%
  \BibitemOpen
  \bibfield  {author} {\bibinfo {author} {\bibfnamefont {E.}~\bibnamefont
  {Komatsu}} \emph {et~al.} (\bibinfo {collaboration} {WMAP}),\ }\href
  {\doibase 10.1088/0067-0049/180/2/330} {\bibfield  {journal} {\bibinfo
  {journal} {Astrophys. J. Suppl.}\ }\textbf {\bibinfo {volume} {180}},\
  \bibinfo {pages} {330} (\bibinfo {year} {2009})},\ \Eprint
  {http://arxiv.org/abs/0803.0547} {arXiv:0803.0547 [astro-ph]} \BibitemShut
  {NoStop}%
\bibitem [{\citenamefont {Ade}\ \emph {et~al.}(2015{\natexlab{a}})\citenamefont
  {Ade} \emph {et~al.}}]{Ade:2015lrj}%
  \BibitemOpen
  \bibfield  {author} {\bibinfo {author} {\bibfnamefont {P.~A.~R.}\
  \bibnamefont {Ade}} \emph {et~al.} (\bibinfo {collaboration} {Planck}),\
  }\href@noop {} {\  (\bibinfo {year} {2015}{\natexlab{a}})},\ \Eprint
  {http://arxiv.org/abs/1502.02114} {arXiv:1502.02114 [astro-ph.CO]}
  \BibitemShut {NoStop}%
\bibitem [{\citenamefont {Ade}\ \emph {et~al.}(2015{\natexlab{b}})\citenamefont
  {Ade} \emph {et~al.}}]{Ade:2015ava}%
  \BibitemOpen
  \bibfield  {author} {\bibinfo {author} {\bibfnamefont {P.~A.~R.}\
  \bibnamefont {Ade}} \emph {et~al.} (\bibinfo {collaboration} {Planck}),\
  }\href@noop {} {\  (\bibinfo {year} {2015}{\natexlab{b}})},\ \Eprint
  {http://arxiv.org/abs/1502.01592} {arXiv:1502.01592 [astro-ph.CO]}
  \BibitemShut {NoStop}%
\bibitem [{\citenamefont {Sievers}\ \emph {et~al.}(2013)\citenamefont {Sievers}
  \emph {et~al.}}]{Sievers:2013ica}%
  \BibitemOpen
  \bibfield  {author} {\bibinfo {author} {\bibfnamefont {J.~L.}\ \bibnamefont
  {Sievers}} \emph {et~al.} (\bibinfo {collaboration} {Atacama Cosmology
  Telescope}),\ }\href {\doibase 10.1088/1475-7516/2013/10/060} {\bibfield
  {journal} {\bibinfo  {journal} {JCAP}\ }\textbf {\bibinfo {volume} {1310}},\
  \bibinfo {pages} {060} (\bibinfo {year} {2013})},\ \Eprint
  {http://arxiv.org/abs/1301.0824} {arXiv:1301.0824 [astro-ph.CO]} \BibitemShut
  {NoStop}%
\bibitem [{\citenamefont {Penrose}(1989)}]{Penrose:1988mg}%
  \BibitemOpen
  \bibfield  {author} {\bibinfo {author} {\bibfnamefont {R.}~\bibnamefont
  {Penrose}},\ }\href {\doibase 10.1111/j.1749-6632.1989.tb50513.x} {\bibfield
  {journal} {\bibinfo  {journal} {Annals N.Y.Acad.Sci.}\ }\textbf {\bibinfo
  {volume} {571}},\ \bibinfo {pages} {249} (\bibinfo {year}
  {1989})}\BibitemShut {NoStop}%
\bibitem [{\citenamefont {Gibbons}\ and\ \citenamefont
  {Turok}(2008)}]{Gibbons:2006pa}%
  \BibitemOpen
  \bibfield  {author} {\bibinfo {author} {\bibfnamefont {G.}~\bibnamefont
  {Gibbons}}\ and\ \bibinfo {author} {\bibfnamefont {N.}~\bibnamefont
  {Turok}},\ }\href {\doibase 10.1103/PhysRevD.77.063516} {\bibfield  {journal}
  {\bibinfo  {journal} {Phys.Rev.}\ }\textbf {\bibinfo {volume} {D77}},\
  \bibinfo {pages} {063516} (\bibinfo {year} {2008})},\ \Eprint
  {http://arxiv.org/abs/hep-th/0609095} {arXiv:hep-th/0609095 [hep-th]}
  \BibitemShut {NoStop}%
\bibitem [{\citenamefont {Berezhiani}\ and\ \citenamefont
  {Trodden}(2015)}]{Berezhiani:2015ola}%
  \BibitemOpen
  \bibfield  {author} {\bibinfo {author} {\bibfnamefont {L.}~\bibnamefont
  {Berezhiani}}\ and\ \bibinfo {author} {\bibfnamefont {M.}~\bibnamefont
  {Trodden}},\ }\href {\doibase 10.1016/j.physletb.2015.08.007} {\bibfield
  {journal} {\bibinfo  {journal} {Phys. Lett.}\ }\textbf {\bibinfo {volume}
  {B749}},\ \bibinfo {pages} {425} (\bibinfo {year} {2015})},\ \Eprint
  {http://arxiv.org/abs/1504.01730} {arXiv:1504.01730 [hep-th]} \BibitemShut
  {NoStop}%
\bibitem [{\citenamefont {{Steinhardt}}(1983)}]{1983veu..conf..251S}%
  \BibitemOpen
  \bibfield  {author} {\bibinfo {author} {\bibfnamefont {P.~J.}\ \bibnamefont
  {{Steinhardt}}},\ }in\ \href@noop {} {\emph {\bibinfo {booktitle} {Very Early
  Universe}}},\ \bibinfo {editor} {edited by\ \bibinfo {editor} {\bibfnamefont
  {G.~W.}\ \bibnamefont {{Gibbons}}}, \bibinfo {editor} {\bibfnamefont {S.~W.}\
  \bibnamefont {{Hawking}}}, \ and\ \bibinfo {editor} {\bibfnamefont
  {S.~T.~C.}\ \bibnamefont {{Siklos}}}}\ (\bibinfo {year} {1983})\ pp.\
  \bibinfo {pages} {251--266}\BibitemShut {NoStop}%
\bibitem [{\citenamefont {Vilenkin}(1983)}]{Vilenkin:1983xq}%
  \BibitemOpen
  \bibfield  {author} {\bibinfo {author} {\bibfnamefont {A.}~\bibnamefont
  {Vilenkin}},\ }\href {\doibase 10.1103/PhysRevD.27.2848} {\bibfield
  {journal} {\bibinfo  {journal} {Phys.Rev.}\ }\textbf {\bibinfo {volume}
  {D27}},\ \bibinfo {pages} {2848} (\bibinfo {year} {1983})}\BibitemShut
  {NoStop}%
\bibitem [{\citenamefont {Guth}(2000)}]{Guth:2000ka}%
  \BibitemOpen
  \bibfield  {author} {\bibinfo {author} {\bibfnamefont {A.~H.}\ \bibnamefont
  {Guth}},\ }\href {\doibase 10.1016/S0370-1573(00)00037-5} {\bibfield
  {journal} {\bibinfo  {journal} {Phys.Rept.}\ }\textbf {\bibinfo {volume}
  {333}},\ \bibinfo {pages} {555} (\bibinfo {year} {2000})},\ \Eprint
  {http://arxiv.org/abs/astro-ph/0002156} {arXiv:astro-ph/0002156 [astro-ph]}
  \BibitemShut {NoStop}%
\bibitem [{\citenamefont {Guth}(2013)}]{Guth:2013epa}%
  \BibitemOpen
  \bibfield  {author} {\bibinfo {author} {\bibfnamefont {A.~H.}\ \bibnamefont
  {Guth}},\ }in\ \href
  {http://inspirehep.net/record/1275330/files/arXiv:1312.7340.pdf} {\emph
  {\bibinfo {booktitle} {{Proceedings, 25th Solvay Conference on Physics: The
  Theory of the Quantum World}}}}\ (\bibinfo {year} {2013})\ \Eprint
  {http://arxiv.org/abs/1312.7340} {arXiv:1312.7340 [hep-th]} \BibitemShut
  {NoStop}%
\bibitem [{\citenamefont {Guth}\ \emph {et~al.}(2014)\citenamefont {Guth},
  \citenamefont {Kaiser},\ and\ \citenamefont {Nomura}}]{Guth:2013sya}%
  \BibitemOpen
  \bibfield  {author} {\bibinfo {author} {\bibfnamefont {A.~H.}\ \bibnamefont
  {Guth}}, \bibinfo {author} {\bibfnamefont {D.~I.}\ \bibnamefont {Kaiser}}, \
  and\ \bibinfo {author} {\bibfnamefont {Y.}~\bibnamefont {Nomura}},\ }\href
  {\doibase 10.1016/j.physletb.2014.03.020} {\bibfield  {journal} {\bibinfo
  {journal} {Phys.Lett.}\ }\textbf {\bibinfo {volume} {B733}},\ \bibinfo
  {pages} {112} (\bibinfo {year} {2014})},\ \Eprint
  {http://arxiv.org/abs/1312.7619} {arXiv:1312.7619 [astro-ph.CO]} \BibitemShut
  {NoStop}%
\bibitem [{\citenamefont {Linde}(2014)}]{Linde:2014nna}%
  \BibitemOpen
  \bibfield  {author} {\bibinfo {author} {\bibfnamefont {A.}~\bibnamefont
  {Linde}},\ }\href@noop {} {\enquote {\bibinfo {title} {{Inflationary
  Cosmology after Planck 2013}},}\ } (\bibinfo {year} {2014}),\ \Eprint
  {http://arxiv.org/abs/1402.0526} {arXiv:1402.0526 [hep-th]} \BibitemShut
  {NoStop}%
\bibitem [{\citenamefont {Brandenberger}\ and\ \citenamefont
  {Finelli}(2001)}]{Brandenberger:2001bs}%
  \BibitemOpen
  \bibfield  {author} {\bibinfo {author} {\bibfnamefont {R.}~\bibnamefont
  {Brandenberger}}\ and\ \bibinfo {author} {\bibfnamefont {F.}~\bibnamefont
  {Finelli}},\ }\href {\doibase 10.1088/1126-6708/2001/11/056} {\bibfield
  {journal} {\bibinfo  {journal} {JHEP}\ }\textbf {\bibinfo {volume} {0111}},\
  \bibinfo {pages} {056} (\bibinfo {year} {2001})},\ \Eprint
  {http://arxiv.org/abs/hep-th/0109004} {arXiv:hep-th/0109004 [hep-th]}
  \BibitemShut {NoStop}%
\bibitem [{\citenamefont {Lyth}(2002)}]{Lyth:2001pf}%
  \BibitemOpen
  \bibfield  {author} {\bibinfo {author} {\bibfnamefont {D.~H.}\ \bibnamefont
  {Lyth}},\ }\href {\doibase 10.1016/S0370-2693(01)01374-0} {\bibfield
  {journal} {\bibinfo  {journal} {Phys. Lett.}\ }\textbf {\bibinfo {volume}
  {B524}},\ \bibinfo {pages} {1} (\bibinfo {year} {2002})},\ \Eprint
  {http://arxiv.org/abs/hep-ph/0106153} {arXiv:hep-ph/0106153 [hep-ph]}
  \BibitemShut {NoStop}%
\bibitem [{\citenamefont {Hwang}(2002)}]{Hwang:2001ga}%
  \BibitemOpen
  \bibfield  {author} {\bibinfo {author} {\bibfnamefont {J.-c.}\ \bibnamefont
  {Hwang}},\ }\href {\doibase 10.1103/PhysRevD.65.063514} {\bibfield  {journal}
  {\bibinfo  {journal} {Phys. Rev.}\ }\textbf {\bibinfo {volume} {D65}},\
  \bibinfo {pages} {063514} (\bibinfo {year} {2002})},\ \Eprint
  {http://arxiv.org/abs/astro-ph/0109045} {arXiv:astro-ph/0109045 [astro-ph]}
  \BibitemShut {NoStop}%
\bibitem [{\citenamefont {Hwang}\ and\ \citenamefont
  {Noh}(2002{\natexlab{a}})}]{Hwang:2002ks}%
  \BibitemOpen
  \bibfield  {author} {\bibinfo {author} {\bibfnamefont {J.}~\bibnamefont
  {Hwang}}\ and\ \bibinfo {author} {\bibfnamefont {H.}~\bibnamefont {Noh}},\
  }\href {\doibase 10.1016/S0370-2693(02)02598-4} {\bibfield  {journal}
  {\bibinfo  {journal} {Phys. Lett.}\ }\textbf {\bibinfo {volume} {B545}},\
  \bibinfo {pages} {207} (\bibinfo {year} {2002}{\natexlab{a}})},\ \Eprint
  {http://arxiv.org/abs/hep-th/0203193} {arXiv:hep-th/0203193 [hep-th]}
  \BibitemShut {NoStop}%
\bibitem [{\citenamefont {Creminelli}\ \emph {et~al.}(2005)\citenamefont
  {Creminelli}, \citenamefont {Nicolis},\ and\ \citenamefont
  {Zaldarriaga}}]{Creminelli:2004jg}%
  \BibitemOpen
  \bibfield  {author} {\bibinfo {author} {\bibfnamefont {P.}~\bibnamefont
  {Creminelli}}, \bibinfo {author} {\bibfnamefont {A.}~\bibnamefont {Nicolis}},
  \ and\ \bibinfo {author} {\bibfnamefont {M.}~\bibnamefont {Zaldarriaga}},\
  }\href {\doibase 10.1103/PhysRevD.71.063505} {\bibfield  {journal} {\bibinfo
  {journal} {Phys. Rev.}\ }\textbf {\bibinfo {volume} {D71}},\ \bibinfo {pages}
  {063505} (\bibinfo {year} {2005})},\ \Eprint
  {http://arxiv.org/abs/hep-th/0411270} {arXiv:hep-th/0411270 [hep-th]}
  \BibitemShut {NoStop}%
\bibitem [{\citenamefont {Finelli}(2002)}]{Finelli:2002we}%
  \BibitemOpen
  \bibfield  {author} {\bibinfo {author} {\bibfnamefont {F.}~\bibnamefont
  {Finelli}},\ }\href {\doibase 10.1016/S0370-2693(02)02554-6} {\bibfield
  {journal} {\bibinfo  {journal} {Phys. Lett.}\ }\textbf {\bibinfo {volume}
  {B545}},\ \bibinfo {pages} {1} (\bibinfo {year} {2002})},\ \Eprint
  {http://arxiv.org/abs/hep-th/0206112} {arXiv:hep-th/0206112 [hep-th]}
  \BibitemShut {NoStop}%
\bibitem [{\citenamefont {Lehners}\ \emph {et~al.}(2007)\citenamefont
  {Lehners}, \citenamefont {McFadden}, \citenamefont {Turok},\ and\
  \citenamefont {Steinhardt}}]{Lehners:2007ac}%
  \BibitemOpen
  \bibfield  {author} {\bibinfo {author} {\bibfnamefont {J.-L.}\ \bibnamefont
  {Lehners}}, \bibinfo {author} {\bibfnamefont {P.}~\bibnamefont {McFadden}},
  \bibinfo {author} {\bibfnamefont {N.}~\bibnamefont {Turok}}, \ and\ \bibinfo
  {author} {\bibfnamefont {P.~J.}\ \bibnamefont {Steinhardt}},\ }\href
  {\doibase 10.1103/PhysRevD.76.103501} {\bibfield  {journal} {\bibinfo
  {journal} {Phys.Rev.}\ }\textbf {\bibinfo {volume} {D76}},\ \bibinfo {pages}
  {103501} (\bibinfo {year} {2007})},\ \Eprint
  {http://arxiv.org/abs/hep-th/0702153} {arXiv:hep-th/0702153 [HEP-TH]}
  \BibitemShut {NoStop}%
\bibitem [{\citenamefont {Buchbinder}\ \emph
  {et~al.}(2007{\natexlab{a}})\citenamefont {Buchbinder}, \citenamefont
  {Khoury},\ and\ \citenamefont {Ovrut}}]{Buchbinder:2007ad}%
  \BibitemOpen
  \bibfield  {author} {\bibinfo {author} {\bibfnamefont {E.~I.}\ \bibnamefont
  {Buchbinder}}, \bibinfo {author} {\bibfnamefont {J.}~\bibnamefont {Khoury}},
  \ and\ \bibinfo {author} {\bibfnamefont {B.~A.}\ \bibnamefont {Ovrut}},\
  }\href {\doibase 10.1103/PhysRevD.76.123503} {\bibfield  {journal} {\bibinfo
  {journal} {Phys.Rev.}\ }\textbf {\bibinfo {volume} {D76}},\ \bibinfo {pages}
  {123503} (\bibinfo {year} {2007}{\natexlab{a}})},\ \Eprint
  {http://arxiv.org/abs/hep-th/0702154} {arXiv:hep-th/0702154 [hep-th]}
  \BibitemShut {NoStop}%
\bibitem [{\citenamefont {Di~Marco}\ \emph {et~al.}(2003)\citenamefont
  {Di~Marco}, \citenamefont {Finelli},\ and\ \citenamefont
  {Brandenberger}}]{DiMarco:2002eb}%
  \BibitemOpen
  \bibfield  {author} {\bibinfo {author} {\bibfnamefont {F.}~\bibnamefont
  {Di~Marco}}, \bibinfo {author} {\bibfnamefont {F.}~\bibnamefont {Finelli}}, \
  and\ \bibinfo {author} {\bibfnamefont {R.}~\bibnamefont {Brandenberger}},\
  }\href {\doibase 10.1103/PhysRevD.67.063512} {\bibfield  {journal} {\bibinfo
  {journal} {Phys.Rev.}\ }\textbf {\bibinfo {volume} {D67}},\ \bibinfo {pages}
  {063512} (\bibinfo {year} {2003})},\ \Eprint
  {http://arxiv.org/abs/astro-ph/0211276} {arXiv:astro-ph/0211276 [astro-ph]}
  \BibitemShut {NoStop}%
\bibitem [{\citenamefont {Tolley}\ and\ \citenamefont
  {Wesley}(2007)}]{Tolley:2007nq}%
  \BibitemOpen
  \bibfield  {author} {\bibinfo {author} {\bibfnamefont {A.~J.}\ \bibnamefont
  {Tolley}}\ and\ \bibinfo {author} {\bibfnamefont {D.~H.}\ \bibnamefont
  {Wesley}},\ }\href {\doibase 10.1088/1475-7516/2007/05/006} {\bibfield
  {journal} {\bibinfo  {journal} {JCAP}\ }\textbf {\bibinfo {volume} {0705}},\
  \bibinfo {pages} {006} (\bibinfo {year} {2007})},\ \Eprint
  {http://arxiv.org/abs/hep-th/0703101} {arXiv:hep-th/0703101 [hep-th]}
  \BibitemShut {NoStop}%
\bibitem [{\citenamefont {Koyama}\ \emph {et~al.}(2007)\citenamefont {Koyama},
  \citenamefont {Mizuno},\ and\ \citenamefont {Wands}}]{Koyama:2007ag}%
  \BibitemOpen
  \bibfield  {author} {\bibinfo {author} {\bibfnamefont {K.}~\bibnamefont
  {Koyama}}, \bibinfo {author} {\bibfnamefont {S.}~\bibnamefont {Mizuno}}, \
  and\ \bibinfo {author} {\bibfnamefont {D.}~\bibnamefont {Wands}},\ }\href
  {\doibase 10.1088/0264-9381/24/15/010} {\bibfield  {journal} {\bibinfo
  {journal} {Class.Quant.Grav.}\ }\textbf {\bibinfo {volume} {24}},\ \bibinfo
  {pages} {3919} (\bibinfo {year} {2007})},\ \Eprint
  {http://arxiv.org/abs/0704.1152} {arXiv:0704.1152 [hep-th]} \BibitemShut
  {NoStop}%
\bibitem [{\citenamefont {Koyama}\ and\ \citenamefont
  {Wands}(2007)}]{Koyama:2007mg}%
  \BibitemOpen
  \bibfield  {author} {\bibinfo {author} {\bibfnamefont {K.}~\bibnamefont
  {Koyama}}\ and\ \bibinfo {author} {\bibfnamefont {D.}~\bibnamefont {Wands}},\
  }\href {\doibase 10.1088/1475-7516/2007/04/008} {\bibfield  {journal}
  {\bibinfo  {journal} {JCAP}\ }\textbf {\bibinfo {volume} {0704}},\ \bibinfo
  {pages} {008} (\bibinfo {year} {2007})},\ \Eprint
  {http://arxiv.org/abs/hep-th/0703040} {arXiv:hep-th/0703040 [HEP-TH]}
  \BibitemShut {NoStop}%
\bibitem [{\citenamefont {Buchbinder}\ \emph
  {et~al.}(2007{\natexlab{b}})\citenamefont {Buchbinder}, \citenamefont
  {Khoury},\ and\ \citenamefont {Ovrut}}]{Buchbinder:2007tw}%
  \BibitemOpen
  \bibfield  {author} {\bibinfo {author} {\bibfnamefont {E.~I.}\ \bibnamefont
  {Buchbinder}}, \bibinfo {author} {\bibfnamefont {J.}~\bibnamefont {Khoury}},
  \ and\ \bibinfo {author} {\bibfnamefont {B.~A.}\ \bibnamefont {Ovrut}},\
  }\href {\doibase 10.1088/1126-6708/2007/11/076} {\bibfield  {journal}
  {\bibinfo  {journal} {JHEP}\ }\textbf {\bibinfo {volume} {0711}},\ \bibinfo
  {pages} {076} (\bibinfo {year} {2007}{\natexlab{b}})},\ \Eprint
  {http://arxiv.org/abs/0706.3903} {arXiv:0706.3903 [hep-th]} \BibitemShut
  {NoStop}%
\bibitem [{\citenamefont {Hinterbichler}\ and\ \citenamefont
  {Khoury}(2012)}]{Hinterbichler:2011qk}%
  \BibitemOpen
  \bibfield  {author} {\bibinfo {author} {\bibfnamefont {K.}~\bibnamefont
  {Hinterbichler}}\ and\ \bibinfo {author} {\bibfnamefont {J.}~\bibnamefont
  {Khoury}},\ }\href {\doibase 10.1088/1475-7516/2012/04/023} {\bibfield
  {journal} {\bibinfo  {journal} {JCAP}\ }\textbf {\bibinfo {volume} {1204}},\
  \bibinfo {pages} {023} (\bibinfo {year} {2012})},\ \Eprint
  {http://arxiv.org/abs/1106.1428} {arXiv:1106.1428 [hep-th]} \BibitemShut
  {NoStop}%
\bibitem [{\citenamefont {{Li}}(2013)}]{2013PhLB..724..192L}%
  \BibitemOpen
  \bibfield  {author} {\bibinfo {author} {\bibfnamefont {M.}~\bibnamefont
  {{Li}}},\ }\href {\doibase 10.1016/j.physletb.2013.06.035} {\bibfield
  {journal} {\bibinfo  {journal} {Physics Letters B}\ }\textbf {\bibinfo
  {volume} {724}},\ \bibinfo {pages} {192} (\bibinfo {year} {2013})},\ \Eprint
  {http://arxiv.org/abs/1306.0191} {arXiv:1306.0191 [hep-th]} \BibitemShut
  {NoStop}%
\bibitem [{\citenamefont {Li}(2015)}]{Li:2014yla}%
  \BibitemOpen
  \bibfield  {author} {\bibinfo {author} {\bibfnamefont {M.}~\bibnamefont
  {Li}},\ }\href {\doibase 10.1016/j.physletb.2015.01.009} {\bibfield
  {journal} {\bibinfo  {journal} {Phys. Lett.}\ }\textbf {\bibinfo {volume}
  {B741}},\ \bibinfo {pages} {320} (\bibinfo {year} {2015})},\ \Eprint
  {http://arxiv.org/abs/1411.7626} {arXiv:1411.7626 [hep-th]} \BibitemShut
  {NoStop}%
\bibitem [{\citenamefont {Ijjas}\ \emph {et~al.}(2014)\citenamefont {Ijjas},
  \citenamefont {Lehners},\ and\ \citenamefont {Steinhardt}}]{Ijjas:2014fja}%
  \BibitemOpen
  \bibfield  {author} {\bibinfo {author} {\bibfnamefont {A.}~\bibnamefont
  {Ijjas}}, \bibinfo {author} {\bibfnamefont {J.-L.}\ \bibnamefont {Lehners}},
  \ and\ \bibinfo {author} {\bibfnamefont {P.~J.}\ \bibnamefont {Steinhardt}},\
  }\href {\doibase 10.1103/PhysRevD.89.123520} {\bibfield  {journal} {\bibinfo
  {journal} {Phys.Rev.}\ }\textbf {\bibinfo {volume} {D89}},\ \bibinfo {pages}
  {123520} (\bibinfo {year} {2014})},\ \Eprint {http://arxiv.org/abs/1404.1265}
  {arXiv:1404.1265 [astro-ph.CO]} \BibitemShut {NoStop}%
\bibitem [{\citenamefont {Levy}\ \emph {et~al.}(2015)\citenamefont {Levy},
  \citenamefont {Ijjas},\ and\ \citenamefont {Steinhardt}}]{Levy:2015awa}%
  \BibitemOpen
  \bibfield  {author} {\bibinfo {author} {\bibfnamefont {A.~M.}\ \bibnamefont
  {Levy}}, \bibinfo {author} {\bibfnamefont {A.}~\bibnamefont {Ijjas}}, \ and\
  \bibinfo {author} {\bibfnamefont {P.~J.}\ \bibnamefont {Steinhardt}},\ }\href
  {\doibase 10.1103/PhysRevD.92.063524} {\bibfield  {journal} {\bibinfo
  {journal} {Phys. Rev.}\ }\textbf {\bibinfo {volume} {D92}},\ \bibinfo {pages}
  {063524} (\bibinfo {year} {2015})},\ \Eprint
  {http://arxiv.org/abs/1506.01011} {arXiv:1506.01011 [astro-ph.CO]}
  \BibitemShut {NoStop}%
\bibitem [{\citenamefont {Notari}\ and\ \citenamefont
  {Riotto}(2002)}]{Notari:2002yc}%
  \BibitemOpen
  \bibfield  {author} {\bibinfo {author} {\bibfnamefont {A.}~\bibnamefont
  {Notari}}\ and\ \bibinfo {author} {\bibfnamefont {A.}~\bibnamefont
  {Riotto}},\ }\href {\doibase 10.1016/S0550-3213(02)00765-4} {\bibfield
  {journal} {\bibinfo  {journal} {Nucl. Phys.}\ }\textbf {\bibinfo {volume}
  {B644}},\ \bibinfo {pages} {371} (\bibinfo {year} {2002})},\ \Eprint
  {http://arxiv.org/abs/hep-th/0205019} {arXiv:hep-th/0205019 [hep-th]}
  \BibitemShut {NoStop}%
\bibitem [{\citenamefont {Boyle}\ \emph {et~al.}(2004)\citenamefont {Boyle},
  \citenamefont {Steinhardt},\ and\ \citenamefont {Turok}}]{Boyle:2003km}%
  \BibitemOpen
  \bibfield  {author} {\bibinfo {author} {\bibfnamefont {L.~A.}\ \bibnamefont
  {Boyle}}, \bibinfo {author} {\bibfnamefont {P.~J.}\ \bibnamefont
  {Steinhardt}}, \ and\ \bibinfo {author} {\bibfnamefont {N.}~\bibnamefont
  {Turok}},\ }\href {\doibase 10.1103/PhysRevD.69.127302} {\bibfield  {journal}
  {\bibinfo  {journal} {Phys. Rev.}\ }\textbf {\bibinfo {volume} {D69}},\
  \bibinfo {pages} {127302} (\bibinfo {year} {2004})},\ \Eprint
  {http://arxiv.org/abs/hep-th/0307170} {arXiv:hep-th/0307170 [hep-th]}
  \BibitemShut {NoStop}%
\bibitem [{\citenamefont {Fertig}\ \emph {et~al.}(2014)\citenamefont {Fertig},
  \citenamefont {Lehners},\ and\ \citenamefont {Mallwitz}}]{Fertig:2013kwa}%
  \BibitemOpen
  \bibfield  {author} {\bibinfo {author} {\bibfnamefont {A.}~\bibnamefont
  {Fertig}}, \bibinfo {author} {\bibfnamefont {J.-L.}\ \bibnamefont {Lehners}},
  \ and\ \bibinfo {author} {\bibfnamefont {E.}~\bibnamefont {Mallwitz}},\
  }\href {\doibase 10.1103/PhysRevD.89.103537} {\bibfield  {journal} {\bibinfo
  {journal} {Phys. Rev.}\ }\textbf {\bibinfo {volume} {D89}},\ \bibinfo {pages}
  {103537} (\bibinfo {year} {2014})},\ \Eprint {http://arxiv.org/abs/1310.8133}
  {arXiv:1310.8133 [hep-th]} \BibitemShut {NoStop}%
\bibitem [{\citenamefont {Berera}\ and\ \citenamefont
  {Fang}(1995)}]{Berera:1995wh}%
  \BibitemOpen
  \bibfield  {author} {\bibinfo {author} {\bibfnamefont {A.}~\bibnamefont
  {Berera}}\ and\ \bibinfo {author} {\bibfnamefont {L.-Z.}\ \bibnamefont
  {Fang}},\ }\href {\doibase 10.1103/PhysRevLett.74.1912} {\bibfield  {journal}
  {\bibinfo  {journal} {Phys. Rev. Lett.}\ }\textbf {\bibinfo {volume} {74}},\
  \bibinfo {pages} {1912} (\bibinfo {year} {1995})},\ \Eprint
  {http://arxiv.org/abs/astro-ph/9501024} {arXiv:astro-ph/9501024 [astro-ph]}
  \BibitemShut {NoStop}%
\bibitem [{\citenamefont {Berera}(1995)}]{Berera:1995ie}%
  \BibitemOpen
  \bibfield  {author} {\bibinfo {author} {\bibfnamefont {A.}~\bibnamefont
  {Berera}},\ }\href {\doibase 10.1103/PhysRevLett.75.3218} {\bibfield
  {journal} {\bibinfo  {journal} {Phys. Rev. Lett.}\ }\textbf {\bibinfo
  {volume} {75}},\ \bibinfo {pages} {3218} (\bibinfo {year} {1995})},\ \Eprint
  {http://arxiv.org/abs/astro-ph/9509049} {arXiv:astro-ph/9509049 [astro-ph]}
  \BibitemShut {NoStop}%
\bibitem [{\citenamefont {Berera}(1996)}]{Berera:1996nv}%
  \BibitemOpen
  \bibfield  {author} {\bibinfo {author} {\bibfnamefont {A.}~\bibnamefont
  {Berera}},\ }\href {\doibase 10.1103/PhysRevD.54.2519} {\bibfield  {journal}
  {\bibinfo  {journal} {Phys. Rev.}\ }\textbf {\bibinfo {volume} {D54}},\
  \bibinfo {pages} {2519} (\bibinfo {year} {1996})},\ \Eprint
  {http://arxiv.org/abs/hep-th/9601134} {arXiv:hep-th/9601134 [hep-th]}
  \BibitemShut {NoStop}%
\bibitem [{\citenamefont {Berera}(2006)}]{Berera:2006xq}%
  \BibitemOpen
  \bibfield  {author} {\bibinfo {author} {\bibfnamefont {A.}~\bibnamefont
  {Berera}},\ }\href {\doibase 10.1080/00107510500392030} {\bibfield  {journal}
  {\bibinfo  {journal} {Contemp. Phys.}\ }\textbf {\bibinfo {volume} {47}},\
  \bibinfo {pages} {33} (\bibinfo {year} {2006})},\ \Eprint
  {http://arxiv.org/abs/0809.4198} {arXiv:0809.4198 [hep-ph]} \BibitemShut
  {NoStop}%
\bibitem [{\citenamefont {Berera}\ \emph {et~al.}(2009)\citenamefont {Berera},
  \citenamefont {Moss},\ and\ \citenamefont {Ramos}}]{Berera:2008ar}%
  \BibitemOpen
  \bibfield  {author} {\bibinfo {author} {\bibfnamefont {A.}~\bibnamefont
  {Berera}}, \bibinfo {author} {\bibfnamefont {I.~G.}\ \bibnamefont {Moss}}, \
  and\ \bibinfo {author} {\bibfnamefont {R.~O.}\ \bibnamefont {Ramos}},\ }\href
  {\doibase 10.1088/0034-4885/72/2/026901} {\bibfield  {journal} {\bibinfo
  {journal} {Rept. Prog. Phys.}\ }\textbf {\bibinfo {volume} {72}},\ \bibinfo
  {pages} {026901} (\bibinfo {year} {2009})},\ \Eprint
  {http://arxiv.org/abs/0808.1855} {arXiv:0808.1855 [hep-ph]} \BibitemShut
  {NoStop}%
\bibitem [{\citenamefont {Bastero-Gil}\ and\ \citenamefont
  {Berera}(2009)}]{BasteroGil:2009ec}%
  \BibitemOpen
  \bibfield  {author} {\bibinfo {author} {\bibfnamefont {M.}~\bibnamefont
  {Bastero-Gil}}\ and\ \bibinfo {author} {\bibfnamefont {A.}~\bibnamefont
  {Berera}},\ }\href {\doibase 10.1142/S0217751X09044206} {\bibfield  {journal}
  {\bibinfo  {journal} {Int. J. Mod. Phys.}\ }\textbf {\bibinfo {volume}
  {A24}},\ \bibinfo {pages} {2207} (\bibinfo {year} {2009})},\ \Eprint
  {http://arxiv.org/abs/0902.0521} {arXiv:0902.0521 [hep-ph]} \BibitemShut
  {NoStop}%
\bibitem [{\citenamefont {Green}\ \emph {et~al.}(2009)\citenamefont {Green},
  \citenamefont {Horn}, \citenamefont {Senatore},\ and\ \citenamefont
  {Silverstein}}]{Green:2009ds}%
  \BibitemOpen
  \bibfield  {author} {\bibinfo {author} {\bibfnamefont {D.}~\bibnamefont
  {Green}}, \bibinfo {author} {\bibfnamefont {B.}~\bibnamefont {Horn}},
  \bibinfo {author} {\bibfnamefont {L.}~\bibnamefont {Senatore}}, \ and\
  \bibinfo {author} {\bibfnamefont {E.}~\bibnamefont {Silverstein}},\ }\href
  {\doibase 10.1103/PhysRevD.80.063533} {\bibfield  {journal} {\bibinfo
  {journal} {Phys. Rev.}\ }\textbf {\bibinfo {volume} {D80}},\ \bibinfo {pages}
  {063533} (\bibinfo {year} {2009})},\ \Eprint {http://arxiv.org/abs/0902.1006}
  {arXiv:0902.1006 [hep-th]} \BibitemShut {NoStop}%
\bibitem [{\citenamefont {Calzetta}\ and\ \citenamefont
  {Hu}(2008)}]{calzetta2008nonequilibrium}%
  \BibitemOpen
  \bibfield  {author} {\bibinfo {author} {\bibfnamefont {E.}~\bibnamefont
  {Calzetta}}\ and\ \bibinfo {author} {\bibfnamefont {B.}~\bibnamefont {Hu}},\
  }\href {https://books.google.com/books?id=BRJ7ryt2l1IC} {\emph {\bibinfo
  {title} {Nonequilibrium Quantum Field Theory}}},\ Cambridge Monographs on
  Mathematical Physics\ (\bibinfo  {publisher} {Cambridge University Press},\
  \bibinfo {year} {2008})\BibitemShut {NoStop}%
\bibitem [{\citenamefont {Bastero-Gil}\ \emph {et~al.}(2014)\citenamefont
  {Bastero-Gil}, \citenamefont {Berera}, \citenamefont {Moss},\ and\
  \citenamefont {Ramos}}]{Bastero-Gil:2014jsa}%
  \BibitemOpen
  \bibfield  {author} {\bibinfo {author} {\bibfnamefont {M.}~\bibnamefont
  {Bastero-Gil}}, \bibinfo {author} {\bibfnamefont {A.}~\bibnamefont {Berera}},
  \bibinfo {author} {\bibfnamefont {I.~G.}\ \bibnamefont {Moss}}, \ and\
  \bibinfo {author} {\bibfnamefont {R.~O.}\ \bibnamefont {Ramos}},\ }\href
  {\doibase 10.1088/1475-7516/2014/05/004} {\bibfield  {journal} {\bibinfo
  {journal} {JCAP}\ }\textbf {\bibinfo {volume} {1405}},\ \bibinfo {pages}
  {004} (\bibinfo {year} {2014})},\ \Eprint {http://arxiv.org/abs/1401.1149}
  {arXiv:1401.1149 [astro-ph.CO]} \BibitemShut {NoStop}%
\bibitem [{\citenamefont {Berera}\ \emph {et~al.}(1999)\citenamefont {Berera},
  \citenamefont {Gleiser},\ and\ \citenamefont {Ramos}}]{Berera:1998px}%
  \BibitemOpen
  \bibfield  {author} {\bibinfo {author} {\bibfnamefont {A.}~\bibnamefont
  {Berera}}, \bibinfo {author} {\bibfnamefont {M.}~\bibnamefont {Gleiser}}, \
  and\ \bibinfo {author} {\bibfnamefont {R.~O.}\ \bibnamefont {Ramos}},\ }\href
  {\doibase 10.1103/PhysRevLett.83.264} {\bibfield  {journal} {\bibinfo
  {journal} {Phys. Rev. Lett.}\ }\textbf {\bibinfo {volume} {83}},\ \bibinfo
  {pages} {264} (\bibinfo {year} {1999})},\ \Eprint
  {http://arxiv.org/abs/hep-ph/9809583} {arXiv:hep-ph/9809583 [hep-ph]}
  \BibitemShut {NoStop}%
\bibitem [{\citenamefont {Berera}\ \emph {et~al.}(1998)\citenamefont {Berera},
  \citenamefont {Gleiser},\ and\ \citenamefont {Ramos}}]{Berera:1998gx}%
  \BibitemOpen
  \bibfield  {author} {\bibinfo {author} {\bibfnamefont {A.}~\bibnamefont
  {Berera}}, \bibinfo {author} {\bibfnamefont {M.}~\bibnamefont {Gleiser}}, \
  and\ \bibinfo {author} {\bibfnamefont {R.~O.}\ \bibnamefont {Ramos}},\ }\href
  {\doibase 10.1103/PhysRevD.58.123508} {\bibfield  {journal} {\bibinfo
  {journal} {Phys. Rev.}\ }\textbf {\bibinfo {volume} {D58}},\ \bibinfo {pages}
  {123508} (\bibinfo {year} {1998})},\ \Eprint
  {http://arxiv.org/abs/hep-ph/9803394} {arXiv:hep-ph/9803394 [hep-ph]}
  \BibitemShut {NoStop}%
\bibitem [{\citenamefont {Bartrum}\ \emph {et~al.}(2014)\citenamefont
  {Bartrum}, \citenamefont {Bastero-Gil}, \citenamefont {Berera}, \citenamefont
  {Cerezo}, \citenamefont {Ramos},\ and\ \citenamefont
  {Rosa}}]{Bartrum:2013fia}%
  \BibitemOpen
  \bibfield  {author} {\bibinfo {author} {\bibfnamefont {S.}~\bibnamefont
  {Bartrum}}, \bibinfo {author} {\bibfnamefont {M.}~\bibnamefont
  {Bastero-Gil}}, \bibinfo {author} {\bibfnamefont {A.}~\bibnamefont {Berera}},
  \bibinfo {author} {\bibfnamefont {R.}~\bibnamefont {Cerezo}}, \bibinfo
  {author} {\bibfnamefont {R.~O.}\ \bibnamefont {Ramos}}, \ and\ \bibinfo
  {author} {\bibfnamefont {J.~G.}\ \bibnamefont {Rosa}},\ }\href {\doibase
  10.1016/j.physletb.2014.03.029} {\bibfield  {journal} {\bibinfo  {journal}
  {Phys. Lett.}\ }\textbf {\bibinfo {volume} {B732}},\ \bibinfo {pages} {116}
  (\bibinfo {year} {2014})},\ \Eprint {http://arxiv.org/abs/1307.5868}
  {arXiv:1307.5868 [hep-ph]} \BibitemShut {NoStop}%
\bibitem [{\citenamefont {Yokoyama}\ and\ \citenamefont
  {Linde}(1999)}]{Yokoyama:1998ju}%
  \BibitemOpen
  \bibfield  {author} {\bibinfo {author} {\bibfnamefont {J.}~\bibnamefont
  {Yokoyama}}\ and\ \bibinfo {author} {\bibfnamefont {A.~D.}\ \bibnamefont
  {Linde}},\ }\href {\doibase 10.1103/PhysRevD.60.083509} {\bibfield  {journal}
  {\bibinfo  {journal} {Phys. Rev.}\ }\textbf {\bibinfo {volume} {D60}},\
  \bibinfo {pages} {083509} (\bibinfo {year} {1999})},\ \Eprint
  {http://arxiv.org/abs/hep-ph/9809409} {arXiv:hep-ph/9809409 [hep-ph]}
  \BibitemShut {NoStop}%
\bibitem [{\citenamefont {Hwang}\ and\ \citenamefont
  {Noh}(2002{\natexlab{b}})}]{Hwang:2001fb}%
  \BibitemOpen
  \bibfield  {author} {\bibinfo {author} {\bibfnamefont {J.-c.}\ \bibnamefont
  {Hwang}}\ and\ \bibinfo {author} {\bibfnamefont {H.}~\bibnamefont {Noh}},\
  }\href {\doibase 10.1088/0264-9381/19/3/308} {\bibfield  {journal} {\bibinfo
  {journal} {Class. Quant. Grav.}\ }\textbf {\bibinfo {volume} {19}},\ \bibinfo
  {pages} {527} (\bibinfo {year} {2002}{\natexlab{b}})},\ \Eprint
  {http://arxiv.org/abs/astro-ph/0103244} {arXiv:astro-ph/0103244 [astro-ph]}
  \BibitemShut {NoStop}%
\bibitem [{\citenamefont {Malik}\ and\ \citenamefont
  {Wands}(2005)}]{Malik:2004tf}%
  \BibitemOpen
  \bibfield  {author} {\bibinfo {author} {\bibfnamefont {K.~A.}\ \bibnamefont
  {Malik}}\ and\ \bibinfo {author} {\bibfnamefont {D.}~\bibnamefont {Wands}},\
  }\href {\doibase 10.1088/1475-7516/2005/02/007} {\bibfield  {journal}
  {\bibinfo  {journal} {JCAP}\ }\textbf {\bibinfo {volume} {0502}},\ \bibinfo
  {pages} {007} (\bibinfo {year} {2005})},\ \Eprint
  {http://arxiv.org/abs/astro-ph/0411703} {arXiv:astro-ph/0411703 [astro-ph]}
  \BibitemShut {NoStop}%
\end{thebibliography}%
\bibstyle{natbib} 
\end{document}